\documentclass[11pt]{article}
\usepackage{jheppub}
\usepackage[utf8x]{inputenc}
\pdfoutput=1
\usepackage{graphicx}
\usepackage{subfigure}
\usepackage{amsmath}
\usepackage{bm} 
\usepackage{xspace}
\usepackage{relsize}
\usepackage{placeins}
\usepackage{multirow}
\usepackage{soul}
\usepackage{ulem}

\newcommand{\nc}{\newcommand}

\usepackage{slashed}

\nc{\beq}{\begin{equation}}
\nc{\eeq}{\end{equation}}
\nc{\beqa}{\begin{eqnarray}}
\nc{\eeqa}{\end{eqnarray}}
\nc{\bea}{\begin{eqnarray}}
\nc{\eea}{\end{eqnarray}}
\nc{\barray}{\begin{eqnarray}}
\nc{\earray}{\end{eqnarray}}
\nc{\barrayn}{\begin{eqnarray*}}
\nc{\earrayn}{\end{eqnarray*}}
\nc{\ra}{\rightarrow}
\newcommand{\lsim}{\!\mathrel{\hbox{\rlap{\lower.55ex \hbox{$\sim$}} \kern-.34em \raise.4ex \hbox{$<$}}}}
\newcommand{\gsim}{\!\mathrel{\hbox{\rlap{\lower.55ex \hbox{$\sim$}} \kern-.34em \raise.4ex \hbox{$>$}}}}
\nc{\Tr}{{\rm Tr}}
\nc{\slsh}{\slash\hspace*{-0.22cm}}

\def\be{\begin{equation}}
\def\ee{\end{equation}}
\def\bea{\begin{eqnarray}}
\def\eea{\end{eqnarray}}

\nc{\infinity}{\infty}
\nc{\mc}{\mathcal}
\nc{\M}{\mathcal{M}}

\def\to{\rightarrow}

\begin{document}

\title{Dark Matter Freeze-In with a Heavy Mediator: \\Beyond the EFT Approach}

\author[a,b]{Evan Frangipane,}
\author[a,b]{Stefania Gori,}
\author[c,d]{Bibhushan Shakya}

\affiliation[a]{Santa Cruz Institute for Particle Physics, University of California, Santa Cruz, CA 95064, USA}
\affiliation[b]{Department of Physics, 1156 High St., University of California Santa Cruz, Santa Cruz, CA 95064, USA}
\affiliation[c]{Deutsches Elektronen-Synchrotron DESY,
Notkestr. 85, 22607 Hamburg, Germany}
\affiliation[d]{CERN, Theoretical Physics Department, 1211 Geneva 23, Switzerland}

\emailAdd{efrangip@ucsc.edu}
\emailAdd{sgori@ucsc.edu}
\emailAdd{bibhushan.shakya@desy.de}

\preprint{CERN-TH-2021-120\\ \rightline{DESY 21-117}}

\abstract{We study dark matter freeze-in scenarios where the mass of the mediator particle that couples dark matter to  the Standard Model is larger than the reheat temperature, $T_{RH}$, in the early Universe. In such  setups, the standard approach is to work with an effective field theory (EFT) where the mediator is integrated out. We examine the validity of this approach in various generic s- and t-channel mediator frameworks.  We find that the EFT approach breaks down when the mediator  mass is between one to two orders of magnitude  larger than $T_{RH}$ due to various  effects such as s-channel resonance, a small thermally-suppressed abundance of the mediator, or decays of Standard Model particles through loops induced by the mediator. This highlights the necessity of including these contributions in such dark matter freeze-in studies. We also discuss the collider phenomenology of the heavy mediators, which is qualitatively different from standard freeze-in scenarios. We highlight that, due to the low $T_{RH}$, the Standard Model-dark matter coupling in these scenarios can be relatively larger than in standard freeze-in scenarios, improving the testability prospects of these setups. 
}

\maketitle
\section{Introduction and Motivation}

The freeze-in mechanism \cite{McDonald:2001vt,Hall:2009bx} has been extensively studied in the literature as a viable production mechanism for the observed relic abundance of dark matter (DM). In contrast to thermal freeze-out scenarios, the freeze-in mechanism is characterized by couplings between the dark matter particle and the Standard Model (SM) thermal bath that are so feeble that the two populations never thermalize. The dark matter abundance is instead built up gradually over the cosmological history through these feeble interactions. Such feeble couplings suppress most experimental probes of dark matter such as indirect or direct detection. The most promising detection avenues consist of producing dark matter parent particles at high energy colliders and observing their displaced decays, see, e.g.\cite{Co:2015pka,DEramo:2017ecx,Belanger:2018sti,Garny:2018ali,No:2019gvl,Calibbi:2021fld}.

Dark matter freeze-in scenarios can broadly be classified into two categories. If dark matter is produced via renormalizable interactions or decays of heavier particles, production dominantly occurs at temperatures close to the mass of the decaying or annihilating particles (in such cases, obtaining the correct relic density requires extremely small couplings); this scenario is dubbed infrared (IR) freeze-in. In contrast, if higher dimensional operators are involved, dark matter production primarily occurs at the highest temperatures, leading to scenarios referred to as ultraviolet (UV) freeze-in \cite{Elahi:2014fsa}, where the relic abundance is sensitive to the highest temperature reached by the thermal bath, the reheat temperature, $T_{RH}$, at which the radiation dominated evolution of the Universe commences after inflation. Although this dependence on UV physics is unattractive,  such scenarios are inevitable in several beyond the Standard Model (BSM) theories. If these mediator particles are heavier than $T_{RH}$\,\footnote{Recall that there is no experimental evidence at present that $T_{RH}$ was much higher than $\sim$MeV, required for successful Big Bang Nucleosynthesis (BBN).}, integrating them out gives rise to higher dimensional operators in an effective field theory (EFT) relevant for early Universe cosmology, leading to UV rather than IR freeze-in of dark matter \cite{Barman:2020plp}. In such cases, the feeble interactions required to produce the correct dark matter relic density are a natural consequence of integrating out the heavy mediators of mass $M$, which leads to the effective couplings getting suppressed by powers of $ T_{RH}/M$. 

In this paper, we explore various cosmological and collider aspects of UV freeze-in scenarios in the presence of heavy mediator particles with $M_{med}>T_{RH}$. Our focus is to study the limitations of the EFT obtained by integrating out the heavy mediators, and how this affects the relic density calculation.  We find that considerations of the existence of the heavy mediators can give rise to important effects not present in the EFT treatment, such as enhanced dark matter production from resonant effects, modification of the DM momentum distribution, and loop induced decays of SM particles into DM. In this paper, we will explore such scenarios in frameworks with s- as well as t-channel mediators. For s-channel mediator scenarios, we will consider scalar and vector mediators that mix with the SM Higgs and $Z$ boson respectively, as well as a heavy $Z'$ that couples directly to both SM and dark matter. For the t-channel mediator scenario, we will consider a new heavy scalar that couples to SM fermions and the DM candidate. Several papers in the literature have studied DM freeze-in with s-channel scalar \cite{Klasen:2013ypa,Merle:2013wta,Roland:2014vba,Merle:2015oja,Roland:2015yoa,Konig:2016dzg,Roland:2016gli,Darme:2019wpd}, vector \cite{Chu:2013jja,Gehrlein:2019iwl,Bhattacharyya:2018evo}, as well as spin-2 \cite{Bernal:2018qlk} mediators. Likewise, t-channel mediators have been discussed in \cite{Belanger:2018sti,Covi:2002vw}. For more general studies of freeze-in via portal mediators, also see \cite{Blennow:2013jba,Mambrini:2013iaa,Chu:2011be}. Most of these studies consider reheat temperatures above the mediator mass, so that the mediator is part of the thermal bath. Our paper, which studies the opposite regime, is therefore complementary to these studies. It is interesting to note that our scenario interpolates between the standard UV freeze-in scenario ($T_{RH}\!\ll\!M_{med}$), and the standard IR freeze-in scenario ($T_{RH} > M_{med}$).

In this paper, we also study the collider phenomenology of such mediators. In general, collider-accessible particles that can produce the correct dark matter abundance via IR freeze-in have very long lifetimes due to the feeble couplings involved, resulting in decays outside collider detectors. Therefore, IR freeze-in scenarios generically require non-standard cosmological histories in order to obtain mediators with modified lifetimes that can be directly tested at colliders, as well as larger production cross sections \cite{Co:2015pka,DEramo:2017ecx,Belanger:2018sti,Garny:2018ali,No:2019gvl,Calibbi:2021fld}. This does not apply to 
the frameworks of interest to us, which feature much larger couplings of the heavy mediator with SM and dark matter particles. Hence we find that collider signatures of such setup are qualitatively different from those expected from standard freeze-in scenarios.  

The paper is organized as follows: In Section \ref{sec:overview}, we provide an overview of the simplified models and cosmological history we base our study on. Cosmological aspects of dark matter freeze-in for the cases of s- and t-channel mediators are discussed  in Sections \ref{sec:schannel} and \ref{sec:tchannel}, respectively. Section \ref{sec:pheno} is devoted to a discussion of various phenomenological aspects of such heavy mediator frameworks. We end with a summary of our main findings in Section \ref{sec:discussion}. A brief discussion of the impact of the epoch before radiation domination on dark matter production is presented in Appendix \ref{app:inflaton}. Finally, in Appendix \ref{app:models}, we present some specific model realizations of our s- and t-channel mediated freeze-in scenarios. 
  
\section{Simplified Models}
\label{sec:overview}

We perform our analyses in the framework of simplified models. We consider a Dirac fermion dark matter particle, $X$, with mass $m_X$, that is stable and singlet under the SM gauge group. 
$X$ interacts with the SM fermions through a heavy mediator. We choose SM fermions rather than gauge bosons since they are lighter and therefore more abundant in the thermal bath at low temperatures. 
Integrating out the mediator therefore gives rise to an EFT with four-fermion interactions between a pair of DM particles and a pair of SM fermions, $f\bar{f}\leftrightarrow \bar X X$, which will produce DM via UV freeze-in. We consider this setup under two broad categories: an s-channel (scalar or vector) mediator (Sec.\,\ref{sec:schannelov}) and a t-channel mediator (Sec.\,\ref{Sec:tchannelv}).

\subsection{S-Channel Mediator}
\label{sec:schannelov}

\subsubsection{Scalar mediator}

We consider a Higgs-portal model, consisting of a SM-singlet real scalar mediator, $\hat S$, with the following interactions:
\be
\mathcal{L}\supset -y_S \hat S \bar X X+\mu_S^2\hat S^2-\lambda |H|^2\hat S^2-\lambda_s \hat S^4\,.
\label{eq:lagrangianschannel}
\ee
$\hat S$ can obtain a non-zero vacuum expectation value (VEV), $s$. If so, it mixes with the SM Higgs. Defining $\hat S=(S_s+s)/\sqrt 2$ and $H=\left(\begin{array}{c} 0\\ \frac{h_{\rm{SM}}+v}{\sqrt 2}\end{array}\right)$ with $v=246$ GeV, the SM-like Higgs $h$ and the mass eigenstate $S$ can be written as
\beq
h=\cos\theta_h h_{\rm{SM}}-\sin\theta_h S_s,~~S=\sin\theta_h h_{\rm{SM}}+\cos\theta_h S_s,~~{\rm{with}}~~\sin(2\theta_h)\,= \frac{2\lambda v s}{(m_S^2-m_h^2)}\,,
\eeq
 where $m_S$ and $m_h$ are the $S$ and $h$ masses, respectively. Note that the mixing angle $\theta_h$ can be chosen relatively independently of the value of $m_S$. This mixing gives rise to $f\bar{f}\leftrightarrow \bar X X$ interactions mediated by both $S$ and $h$ in the s-channel. The couplings involved in these interactions are given by
\beq\label{eq:CouplingScalarModel}
\mathcal L_{\rm{DM}}=-y_S  \bar X X (\cos\theta_h S-\sin\theta_h h)-y_f\bar ff (\sin\theta_h S+\cos\theta_h h)\,,
\eeq
where $y_f=m_f/v$ is the SM fermion Yukawa. The DM mass can arise from the Yukawa interaction with $S$, from additional interactions with a broader range of dark sector particles, or be vector-like. We will therefore treat it as a free parameter in our numerical investigations. 

\subsubsection{Vector mediator}\label{Sec:vectorMediator}

\vskip 1mm 
\noindent\textbf{Kinetic Mixing}
\vskip 1mm

A second possibility for the s-channel mediator is a dark $U(1)$ gauge boson (the dark photon \cite{Holdom:1985ag}), $\hat Z^\prime$, that mixes with the SM hypercharge gauge boson, $\hat B$. The corresponding Lagrangian is given by
\be
\mathcal{L}\supset \frac{\epsilon}{2}\hat B^{\mu\nu}\hat Z^\prime_{\mu\nu}+\frac{1}{2}m_{\hat Z^\prime}^2\hat Z^{\prime\mu}\hat Z^{\prime}_\mu+i \bar X\gamma^\mu D_\mu X-m_X X \bar X\,,
\label{eq:lagrangianschannelAp}
\ee
where $D_\mu=\partial_\mu-i g_D q_D\hat Z^\prime_{\mu}$, with $g_D$ and $q_D$ the dark $U(1)$ gauge coupling and charge of the $X$ particle, respectively. The $\hat Z^\prime$ mass, $m_{\hat Z^\prime}$, can come either from the interaction with a dark Higgs or from the Stueckelberg mechanism \cite{Stueckelberg:1938hvi}. The kinetic mixing parameter, $\epsilon$, induces a mixing of the $\hat Z^\prime$ with the SM hypercharge gauge boson. In particular, if we define $Z^{\prime}_0= \sqrt{1-\epsilon^2}~\hat Z^\prime$, the dark $Z^\prime$ and the SM $Z$ mass eigenstates are given by
\beq\label{eq:ZprimeMix}
Z^\prime=-\sin\alpha \hat Z+\cos\alpha Z^{\prime}_0,~~Z=\cos\alpha \hat Z+\sin\alpha Z^{\prime}_0,~~{\rm{with}}~~\tan\alpha\sim \frac{m_Z^2}{m_{\hat Z^\prime}^2}\epsilon\sin\theta_W\,,
\eeq
where $\theta_W$ is the Weinberg angle, $\hat Z$ is the would-be SM $Z$ boson, and the last approximated expression only holds for $\epsilon\ll 1$ and $m_{\hat Z^\prime}\gg m_Z$. In the same limit, the mass of the physical $Z^\prime$ is $m_{\hat Z^\prime}$ (corrections arise at the $\epsilon^2$ order).
The mixing in Eq.\,(\ref{eq:ZprimeMix}) is responsible for the $Z^\prime$ coupling with SM fermions, and for the SM $Z$ boson with DM:
\beq
\mathcal{L}\supset  g^\prime_L Z^\prime_\mu\bar f_L\gamma^\mu f_L \: + \: g^\prime_R Z^\prime_\mu\bar f_R\gamma^\mu f_R \: + \: g_{X}^\prime Z^\prime_\mu \bar X\gamma^\mu X + (Z^\prime\leftrightarrow Z)\,,
\eeq
where the couplings are given by
\begin{align}\label{eq:Zpcoups}
	& g^\prime_L =\frac{g_W}{\cos \theta_W}\left (-\sin \alpha \: (\cos^2 \theta_W  \: T_3 - \sin^2 \theta_W \:  Y_L) + \cos \alpha \: \eta \: \sin \theta_W \: Y_L\right) , \nonumber \\
	& g^\prime_R= \frac{g_W}{\cos \theta_W} \left(-\sin \alpha \: (-\sin^2 \theta_W  \: Y_R ) + \cos \alpha \: \eta \: \sin \theta_W \: Y_R\right) , \\
	& g_{X}^\prime= q_D g_D \cos \alpha, \nonumber
\end{align}
with $\eta\equiv\epsilon/\sqrt{1-\epsilon^2}$. $T_3$ and $Y$ are, respectively, the third component of the isospin and the hypercharge of the SM fermion, where we use the convention $Q=T_3+Y$, with $Q$ the fermion electric charge. The corresponding couplings of the $Z$ boson ($g_L,g_R,g_X$) are obtained via the exchange $-\sin\alpha\to\cos\alpha$ and $\cos\alpha\to\sin\alpha$. These couplings give rise to the $f\bar{f}\leftrightarrow X\bar X$ interactions mediated by both the $Z^\prime$ and the SM $Z$ boson.

\vskip 2mm 
\noindent\textbf{\boldmath Gauged $L_\mu-L_\tau$}
\vskip 1mm
In both of the simplified models discussed above, the s-channel mediator obtains couplings to SM fermions via its mixing with a SM particle ($h$ or $Z$). As a consequence, both the mediator and the particle it mixes with can mediate interactions between SM fermions and dark matter. Here we instead consider a setup where the mediator couples directly to SM fermions without necessarily mixing with any SM particle. Such a scenario is realized, for instance, in models obtained by gauging one of the anomaly-free global symmetries of the SM, such as $L_\mu-L_\tau$ or $B-L$. If DM is charged under this symmetry, it can interact with SM fermions via the new $Z^\prime$ without the SM $Z$ boson featuring in the process. 

The minimal $L_\mu-L_\tau$ DM model is defined by the following Lagrangian:
\begin{equation}
{\mathcal L} \supset q_\ell g^{\prime} \left( \bar{\mu} \gamma_{\alpha} \mu - \bar{\tau} \gamma_{\alpha} \tau + \bar{\nu_{\mu}} \gamma_{\alpha} P_L \nu_{\mu} - \bar{\nu_{\tau}} \gamma_{\alpha} P_L \nu_{\tau} \right)Z^{\prime\alpha}+ q_{\chi}\,  g^{\prime}\, \bar{X}\gamma_{\alpha} X Z^{\prime \alpha} +\frac{1}{2}m_{Z^\prime}^2 Z^{\prime\mu}Z^{\prime}_\mu-m_X X \bar X\,,
\label{model:eq1}
\end{equation}
where $q_\ell$ and $q_\chi$ are free parameters quantifying the charge of the SM leptons and DM under the new $U(1)_{\mu-\tau}$ gauge symmetry, and $g^\prime$ the gauge coupling strength of $U(1)_{\mu-\tau}$. In this model, a kinetic mixing between the new $Z^\prime$ and the SM $Z$ boson is induced via loops of SM taus and muons, which are charged under both $U(1)_{\mu-\tau}$ and the hypercharge $U(1)_Y$. The kinetic mixing coefficient is given by 
\be
\epsilon=\frac{q_\ell g^\prime g_1}{12\pi^2}\log\left(\frac{m_\tau^2}{m_\mu^2}\right)\,,
\label{eq:mutaukinetic}
\ee
where $g_1$ is the hypercharge gauge coupling. We will see later that this tiny mixing is inconsequential for dark matter production and phenomenology.

\subsection{T-Channel Mediator}\label{Sec:tchannelv}
Another possibility is the existence of a t-channel mediator. In this paper, we focus on a scalar mediator, $S_T$, that interacts as
\be
\mathcal{L}\supset m_T^2S_T S_T^\dagger-m_X \bar X X-(y_T S_T f \bar{X} +h.c.)\,,
\label{eq:lagrangiantchannel}
\ee
where $f$ is a SM fermion. We assume that the scalar potential is such that $S_T$ does not get a VEV, and therefore does not mix with the SM Higgs boson. The above coupling gives rise to the $f\bar{f}\leftrightarrow \bar X X$ interactions via $S_T$ in the t-channel. 
Such a setup can be realized, e.g. in supersymmetric frameworks, where $S_T$ is identified as a sfermion and $X$ as the bino or axino (see e.g. \cite{Covi:2002vw}).

Note that $S_T$ cannot be a SM singlet but carries the same charges as $\bar f$. For example, if $f$ is a $SU(2)_L$ doublet, $S_T$ must likewise be a doublet with multiple degrees of freedom. While this would give rise to a greater variety of collider signatures, in this paper we focus on the simpler case where $f$ is a right-handed fermion, so that only one, $SU(2)_L$ singlet, mediator is involved. In particular, we will discuss two cases, (1) $f=e_R$, and (2) $f=t_R$, for which the cosmology is qualitatively different. In both cases, $S_T$ couples to the SM $\gamma,Z$ bosons; in the latter case, $S_T$ also couples to gluons.

\subsection{General Features of the Cosmological History}

In the remainder of the paper, we will explore in detail various aspects of dark matter freeze-in in each of the scenarios outlined above. In this section, we provide a brief overview before delving into the details in the following sections. 

We assume that the early Universe after the end of inflation is radiation dominated, and denote the reheat temperature at the beginning of this era as $T_{RH}$. We assume that the epoch before radiation domination, when the Universe is dominated by the energy density of the decaying inflaton field, contributes negligibly to the dark matter abundance (the conditions for the validity of this assumption are discussed in Appendix \ref{app:inflaton}). Therefore, in all scenarios we are interested in, all of the dark matter is produced by freeze-in processes at temperatures below $T_{RH}$.  

The process common to all (s- or t-channel) scenarios is dark matter production from the annihilation of SM fermions, $f\bar{f}\to X\bar{X}$. The dark matter yield from this process can be calculated as \cite{Chu:2013jja} 
\beq\label{eq:abundanceGeneral}
Y_{f\bar f}=\frac{1}{4(2\pi)^8}\frac{1}{{g_*}^S \sqrt{g_*^{\rho}}}\left(\frac{45}{\pi}\right)^{3/2}\frac{M_{Pl}}{m_X}\int_{m_X/T_{RH}}^\infty\,dx \int_{2m_> /T}^\infty z \,(z^2-4x^2)^{1/2} \,K_1(z) \,dz \,|\mathcal M|^2 \,d\Omega\,,
\eeq
where $z=\sqrt{s}/T,~x=m_X/T$, $m_>\equiv{\rm{Max}}(m_f, m_X)$, $M_{\rm{Pl}}$ is the Planck mass, $d\Omega$ is the integral over the solid angle, $s$ is the Mandelstam variable, $K_1(z)$ is the 1st-order modified Bessel function of the second kind, $g_*^{\rho}$ and $g_*^S$ are the effective numbers of degrees of freedom of the thermal bath for the energy and entropy densities, respectively, and $\mathcal M$ is the spin-averaged matrix element for the process. This equation is applicable to the case of s- or t-channel mediators, as well as to the higher dimensional, effective four-fermion interactions, with appropriate specifications of the matrix element $\mathcal M$. 

In addition to this fermion annihilation process, scenario-specific annihilation and decay processes not captured in the EFT, such as annihilation processes that produce one or two mediator particles in the final state,  can also contribute to the dark matter abundance. 

Here we comment on the production of dark matter through the decay of a mediator, concluding that this process should not be added to the aforementioned fermion annihilation process. In the s-channel case, when the mediator mixes with a SM particle (such as the Higgs or $Z$ bosons), decays of the SM particle population in the bath can in principle also contribute to the dark matter abundance. The contribution from the mediator decay can be estimated as 
\begin{eqnarray}\label{eq:HDecaystoDM}
Y_{\rm decay}&\approx&\frac{45}{(1.66) 2\pi^4}\frac{g_{\rm{SM}}\,M_{Pl} \Gamma_{\rm{SM}\to X\bar X}}{m_{\rm{SM}}^2~ {g_*}^S \sqrt{g_*^{\rho}}}\int^\infty_{\frac{m_{\rm{SM}}}{T_{RH}}} K_1(x) x^3 dx\,\nonumber\\
&\approx&\frac{225}{(1.66) 8\pi^4}\frac{g_{\rm{SM}}\,M_{Pl} \Gamma_{\rm{SM}\to X\bar X}}{m_{\rm{SM}}^2~ {g_*}^S \sqrt{g_*^{\rho}}}\,\Gamma\left[7/2,\,m_{\rm{SM}}/T_{RH}\right]\,.
\end{eqnarray}
Here, $g_{\rm{SM}},m_{\rm{SM}}$ are the number of degrees of freedom and mass of the SM mediator, respectively.  The second approximation holds for $m_{\rm{SM}}\!>\!T_{RH}$, with $\Gamma\left[x,y\right]$ the incomplete Gamma function. 
If $T_{RH}\ll m_{\rm{SM}}$, this abundance is Boltzmann suppressed, as encapsulated in $K_1(x)$. We can have $m_{DM}>T_{RH}$ as long as kinematically accessible in the decay. Likewise, the decay of the heavy (non-SM) mediator particle also contributes, with the contribution given by Eq.\,(\ref{eq:HDecaystoDM}) with appropriate replacements. For all mediator and SM decay contributions, the width of the decaying particle is greater than the Hubble rate for all $T_{RH}$ we consider in this paper. Thus these particles should be thought of as resonances rather than long-lived particles during the cosmological epoch of interest for dark matter production, which nevertheless maintain an ``equilibrium" thermal abundance due to rapid inverse decays of SM states (for s-channel mediators), justifying the use of Eq.\,(\ref{eq:HDecaystoDM}) for such scenarios. 

Note, however, that the s-channel $2\to2$ process $f\bar{f}\to X\bar{X}$ contains a resonance regime, where the center of mass energy of the incoming particles matches the mass of the mediator, resulting in an enhancement of the cross section. This resonance regime corresponds to the intermediate particle being produced on shell, then decaying into dark matter particles. This is precisely 
the decay contribution given by Eq.\,(\ref{eq:HDecaystoDM}). Therefore, including both contributions amounts to double counting. In such scenarios, it is therefore sufficient to only consider the annihilation contribution in Eq.\,(\ref{eq:abundanceGeneral}), which includes both on- and off-shell contributions from the mediator (see e.g.\,\cite{Giudice:2003jh,Belanger:2018ccd} for related discussions). Nevertheless, in scenarios where the resonant regime dominates the production process, the decay contribution calculated from Eq.\,(\ref{eq:HDecaystoDM}) will match the yield from Eq.\,(\ref{eq:abundanceGeneral}). In this case, the decay contribution, which is simpler to calculate, can be used to estimate the dark matter abundance. 


\section{Dark Matter Freeze-In: S-Channel Mediator}
\label{sec:schannel}

In this section, we study freeze-in scenarios mediated by s-channel mediators, exploring the interplay between the various production channels in the dark scalar (Sec.\,\ref{Sec:sChannelScalar}) and dark vector (Sec.\,\ref{Sec:sChannelGauge}) cases.  

\subsection{Scalar Mediator}\label{Sec:sChannelScalar}

\subsubsection{Contributions to Dark Matter Freeze-In}

The dark matter abundance receives contributions from several $f\bar f\to X\bar X$ processes. These freeze-in processes have been considered in a EFT framework after integrating out the heavy mediator \cite{Elahi:2014fsa}. While $hh,SS\to X\bar X$ annihilations also contribute, these contributions are negligible for low reheat temperatures $T_{RH}\ll m_h$ as the DM yield is proportional to the square of the abundance of the Higgs or the scalar $S$, which are Boltzmann suppressed.  

As we will show, among the SM fermion annihilations, the SM Higgs exchange processes dominate since we consider the regime $m_h\ll m_S$. 
The identity of the SM fermion that dominates the process depends on $m_X$ and $T_{RH}$. The fermions that couple appreciably to the Higgs and have (relatively) unsuppressed thermal abundance at $T_{RH}\ll m_h$ are the bottom and charm quarks, and $\tau$ leptons. 

The relevant part of the Lagrangian, before integrating out the SM Higgs boson and scalar $S$, is given by
\beq
\mathcal L\supset y_{hXX} h X\bar X + y_{hff} h \bar ff +y_{SXX} S X\bar X + y_{Sff} S \bar ff+ h.c.\,,
\eeq
where the two couplings can be expressed in terms of the Lagrangian parameters and Higgs mixing angle in Eq.\,(\ref{eq:CouplingScalarModel}) as $y_{hXX}=y_S\sin\theta_h,~y_{hff}=-y_f\cos\theta_h,$ and analogously $y_{SXX}=-y_S\cos\theta_h,~y_{Sff}=-y_f\sin\theta_h$.

In the EFT framework, we can compute the DM yield from freeze-in via a four-fermion dimension-6 operator $\frac{1}{\Lambda^2} (f\bar{f})(X \bar X)$, where $\frac{1}{\Lambda^2}=\frac{1}{\Lambda_h^2}+\frac{1}{\Lambda_S^2}$, with $\Lambda_{h,S}=\frac{m_{h,S}}{\sqrt{y_{hff} \,y_{hXX}}}$\footnote{In our numerical calculations, we include the running of the Yukawa coupling, $y_{hff}=y_{hff}(\sqrt s)$.}, as \cite{Elahi:2014fsa}
\beq
Y^{\rm{EFT}}\sim\frac{45\,N_c}{(1.66)\pi^{7}~{g_*}^S \sqrt{g_*^{\rho}}} \frac{M_{Pl} T_{RH}^3}{\Lambda^4}\,,
\label{Yefts}
\eeq 
 where $\text{N}_c$ is the number of colors of the fermion $f$, and we take ${g_*}^S=g_*^{\rho}=100$ as the effective number of relativistic degrees of freedom around the GeV scale. 
Note that this formula assumes that the mass of the annihilating fermion is negligible compared to $T_{RH}$. 

\begin{figure}[t]
\begin{center}
  \includegraphics[width=1\linewidth]{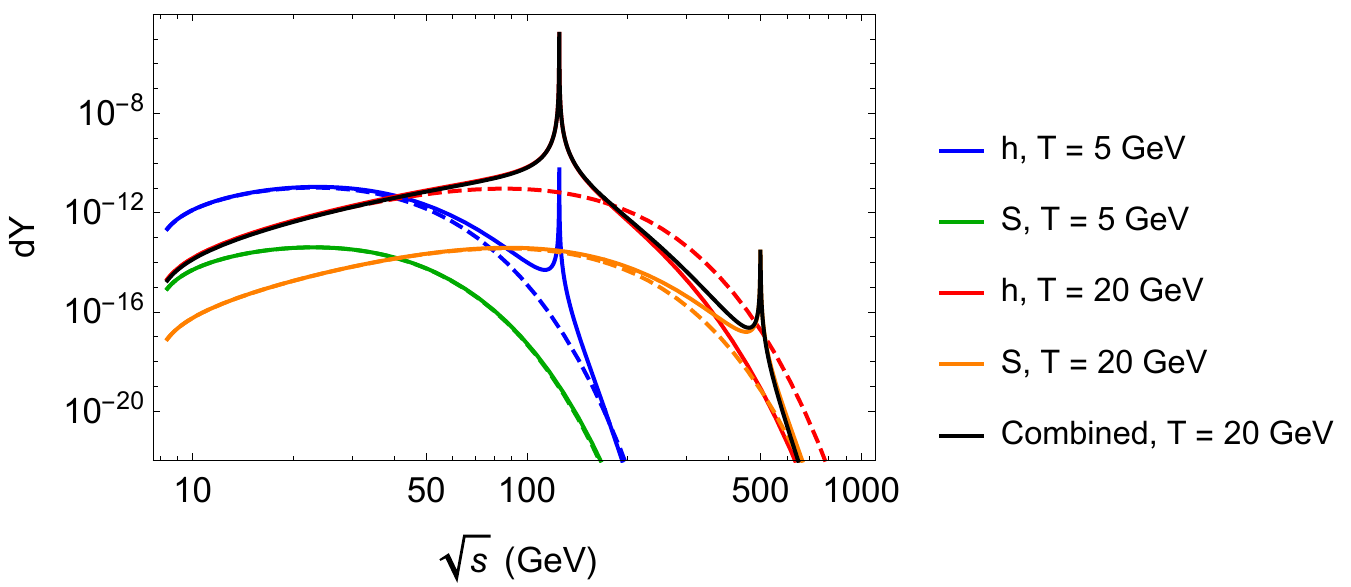}
  \end{center}
  \caption{Dark matter differential yield from the annihilation process $\bar bb\to X\bar X$ as a function of $\sqrt s$, as obtained using the approximate EFT formula (dashed curves) and the full calculation (solid curves) for two temperatures $T=5$ GeV and $T=20$ GeV. We show the SM Higgs ($h$) and heavy scalar ($S$) mediated contributions separately to highlight their relative sizes (the interference term is not shown); the total contribution (including interference) is shown with a black curve. For this plot, we use  $m_S=500$ GeV, $m_X=2$ GeV, $y_{hXX}=10^{-6}$, and $\sin\theta_h=0.1$.}
\label{fig:resonance}
\end{figure}

We now turn to a treatment of this process that takes into account the physical nature of the mediator rather than treating the interaction as an EFT. The full matrix elements in the annihilation process are given by
\beq
|\mathcal M_{2\to 2}|^2=|\mathcal M_{S}|^2+|\mathcal M_{h}|^2+(\mathcal M_{h}\mathcal M^*_{S}+{\rm{h.c.}})\,,
\eeq
where 
\begin{eqnarray}\label{eq:matrixelement}
{|\mathcal M_h|}^2 &=& \frac{4\, \text{N}_c\, y_{hXX}^2\, y_{hff}^2}{(m_h^2-s)^2 + \Gamma_h^2 m_h^2} \frac{s - 4 m_f^2}{2} \frac{s-4 m_X^2}{2},\\\nonumber
	{|\mathcal M_S|}^2 &=& \frac{4\, \text{N}_c\, y_{SXX}^2\, y_{Sff}^2}{(m_S^2-s)^2 + \Gamma_S^2 m_S^2} \frac{s - 4 m_f^2}{2} \frac{s-4 m_X^2}{2}\,,\\\nonumber
	\mathcal M_h\mathcal{M}^*_S+h.c. &=& -8\, \text{N}_c\, (y_{hXX}\, y_{hff}\,)^2 \frac{s^2-(m_h^2+m_S^2)s + m_h^2 m_S^2 + m_h \Gamma_h m_S \Gamma_S}{[(m_h^2-s)^2 + \Gamma_h^2 m_h^2] [(m_S^2-s)^2 + \Gamma_S^2 m_S^2] } \frac{s - 4 m_f^2}{2} \frac{s-4 m_X^2}{2}.
\end{eqnarray}
Here $\Gamma_h$ ($=4.1$ MeV) and $\Gamma_S$ are the widths of the SM Higgs and heavy scalar, respectively. $\Gamma_S=\Gamma^{\rm{SM}}_S\sin^2\theta_h+\Gamma_S^{XX}\cos^2\theta_h$, where $\Gamma^{\rm{SM}}_S$ is the width of the corresponding SM-like Higgs at the same mass, and $\Gamma_S^{XX}$ the width into DM particles. Substituting these matrix elements into Eq.\,(\ref{eq:abundanceGeneral}) enables us to calculate the DM abundance from these annihilation processes. Since the combination of couplings entering the Higgs and $S$ contributions are the same ($|y_{hXX}\, y_{hff}|=|y_{SXX}\, y_{Sff}|$), and the mixing angle $\theta_h$ is treated as a parameter independent of the mass $m_S$, we expect the Higgs exchange to generically dominate over the $S$ exchange or interference term. This is confirmed by Fig.\,\ref{fig:resonance}.

In Fig.\,\ref{fig:resonance}, the solid curves show the differential DM yield contributions from the two mediators, $h$ and $S$, from the $\bar bb\to X\bar X$ process as a function of the center of mass energy at two different temperatures, $T=5$ GeV (in blue and green) and $T=20$ GeV (in red and orange). We choose a benchmark scenario with $m_S=500$ GeV, $m_X=2$ GeV, $y_{hXX}=10^{-6}$, and $\sin\theta_h=0.1$. The Higgs exchange contribution (in blue and red) dominates at most energies, but at higher temperatures the $S$ resonance can produce the dominant effect when $\sqrt s\sim m_S$, as can be seen from the orange curve for $T=20$ GeV. 
For comparison, we also show (as dashed curves) the contributions obtained by dropping the $s$ dependence in the denominators of the matrix elements in Eq.\,(\ref{eq:matrixelement}), which leads to the EFT approximation in Eq.\,(\ref{Yefts}) for $T_{RH}\gg m_b$. This highlights that the approximate EFT results are generally very close to the results of the full calculation, but, as expected, completely miss the resonant regions, which is a crucial aspect of considering the physical nature of the heavy mediator rather than treating the annihilation in an EFT framework. Such resonant contributions grow with $T$, as a larger part of the thermal distributions of the SM fermions can reach the resonant regime.

\subsubsection{Cosmological History}\label{Sec:CosmologyScalarModel}

We now study the interplay between various contributions in producing the measured dark matter relic abundance. 

\begin{figure}[t]
\begin{center}
  \includegraphics[width=1\linewidth]{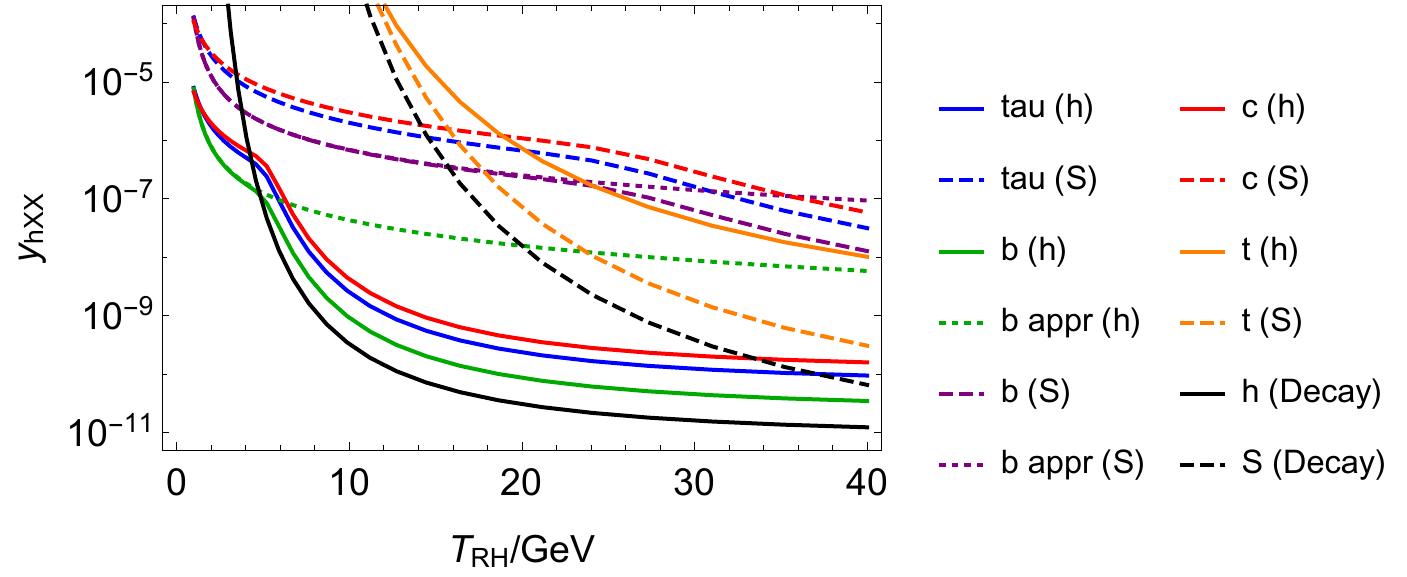}
  \end{center}
\caption{Relative contributions of various annihilation channels, $f\bar f\to X\bar X$, as a function of the reheat temperature for $m_X=1$ GeV, $m_S=500$ GeV, $\sin\theta_h=0.1$. For each channel, we plot the effective Yukawa coupling, $y_{hXX}$, needed to obtain the measured relic abundance. Lower curves correspond to the more dominant channels. The dotted curves are obtained using the approximate EFT formulae. For illustrative purposes, we also include the contributions from $h$ and $S$ decays (black solid and dashed curves, respectively).}
\label{fig:relativecontributions1d}
\end{figure}

In Fig.\,\ref{fig:relativecontributions1d}, we show the relative sizes of the several fermion annihilation channels for a representative choice of heavy scalar mass ($m_S=500$ GeV), heavy scalar-Higgs mixing ($\sin\theta_h=0.1$), and dark matter mass ($m_X=1$ GeV). The y-axis shows the size of the Higgs-DM coupling, $y_{hXX}$, needed for a particular channel to fully provide the observed relic density of dark matter as a function of the reheat temperature. The lower a curve, the more efficient the channel is in producing dark matter. Therefore, the lowest curve represents the most dominant contribution. The solid and dashed curves show the full calculation of the annihilation contributions  mediated by the Higgs and $S$, respectively. We have checked that interference effects are unimportant.
For comparison, the dotted curves show the contributions as computed from the EFT. For illustrative purposes, we also include the contributions from $h$ and $S$ decays (black solid and dashed curves, respectively). 

As we can see, the dominant contribution comes from $b\bar b\to X\bar X$ through the exchange of a Higgs (solid green curve). For very low reheat temperatures, $T_{RH}\lsim 5$ GeV, the EFT calculations match the full calculation (see solid vs. dashed green curves), hence the EFT language appropriately captures the dark matter production. In this regime, the correct dark matter abundance is obtained for $y_{hXX}\sim 10^{-7}-10^{-5}$. On the other hand, for $T_{RH}\gtrsim5$ GeV, the curves depart from the EFT results as the resonant behavior due to the presence of the physical s-channel mediator (SM Higgs) becomes relevant (see also Fig.\,\ref{fig:resonance}). In this region, the value of $y_{hXX}$ required to produce the observed DM relic abundance can be more than two orders of magnitude smaller than that predicted from the EFT treatment.  In this regime, $T_{RH}$ is sufficiently high that production is dominated by the contribution from the resonance region $\sqrt{s}\simeq m_h$, which is effectively captured by calculating the contribution from $h$ decay (solid black curve in the figure, as obtained from Eq.\,(\ref{eq:HDecaystoDM}) with $\Gamma_{{\rm{SM}}\to X\bar X}\to\Gamma_{h\to X\bar X}$ and $m_{\rm{SM}}\to m_h$). Here, we find that couplings as small as $y_{hXX}\sim10^{-11}$ (for $T_{RH}\gsim 20$ GeV) can produce the correct dark matter relic abundance. Such numbers are comparable to the feeble couplings associated with traditional IR dominated freeze-in scenarios.  

These patterns also hold for the contributions from the heavy Higgs exchange\footnote{For simplicity, in the figure we do not include the $hh\to S^*\to X\bar X$ channel, which is justified if the dimensionful coupling $hhS$ is much smaller than $v$. This channel can be larger than the $t\bar t\to S^{(*)}\to X\bar X$ channel but will remain negligible compared to $b\bar b\to h^{(*)}\to X\bar X$.}.
While the total contribution from $S$-mediated processes is always subdominant to that from the SM Higgs at these low $T_{RH}$, it is interesting to note that  $t\bar t\to S^*\to X\bar X$ is more efficient than $t\bar t\to h^*\to X\bar X$ (see the solid vs. dashed orange curves) as the latter does not get any resonant enhancement. 

\begin{figure}[t]
\begin{center}
   \includegraphics[width=0.7\linewidth]{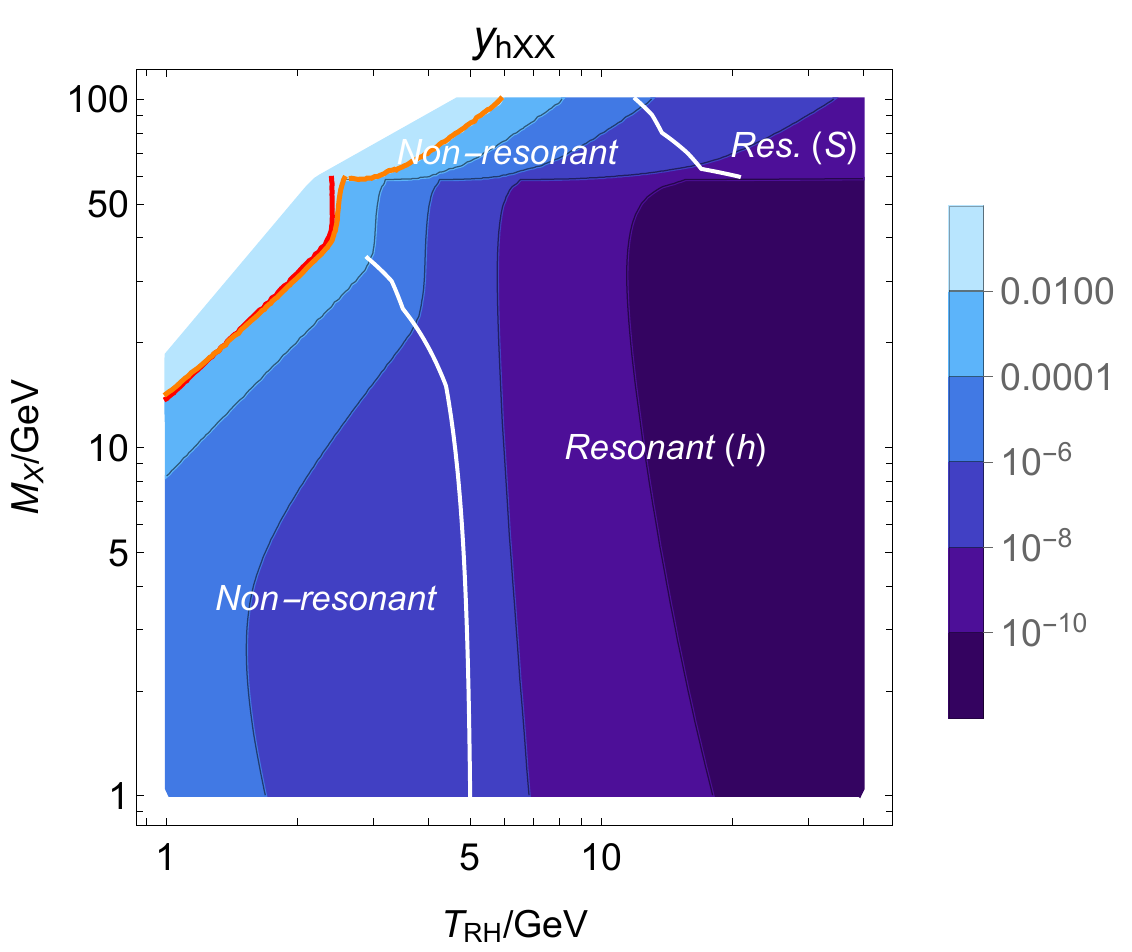}
  \end{center}
	\caption{Contours of the Higgs-DM coupling ($y_{hXX}$) needed to produce the measured DM relic abundance, as a function of the reheat temperature and of the DM mass for $m_S=500$ GeV and $\sin\theta_h=0.1$. The white region corresponds to $y_{hXX}\!>\!1$. The white curves separate regions of parameter space where different contributions dominate dark matter production, as specified by the labels. The red curve represents the LHC bound on Higgs invisible decays (see Sec.\,\ref{Sec:LHCScalar}), whereas the orange curve represents the constraints from current direct detection data from XENON1T \cite{XENON:2018voc} (see Sec.\,\ref{subsec:directdetection}). }
\label{fig:relativecontributions2d}
\end{figure}

In Fig.\,\ref{fig:relativecontributions2d}, we show contours of the Higgs-DM coupling $y_{hXX}$ that produce the measured DM relic abundance as a function of $T_{RH}$ and $m_X$. The sharp features at $m_X\sim m_h/2$ are due to the fact that much larger couplings are needed at higher DM masses as the Higgs resonant enhancement is no longer possible. Relatively large Higgs-DM couplings are needed ($y_{hXX}\sim \mathcal O(0.01)$ and above\footnote{Even for such large couplings, dark matter does not thermalize with the SM bath in these regions of parameter space as $m_h\gg T_{RH}$.}) for $m_{X}\gtrsim\mathcal{O}(10$ GeV) and $T_{RH}\sim\mathcal{O}($GeV). We also show the boundary between regions where dark matter production is dominated by non-resonant annihilation (where the EFT approach provides a good approximation of the yield), and regions where production is dominated by resonant annihilation (where either the $h$ or $S$ decay approximation from Eq.\,(\ref{eq:HDecaystoDM}) appropriately captures the dark matter yield). In the figure, we also show the bound from LHC searches for the Higgs decaying invisibly (red curve) and the direct detection bounds from XENON1T \cite{XENON:2018voc} (orange curve), see Secs.\,\ref{Sec:LHCScalar}, \ref{subsec:directdetection} for more details. These searches already probe part of the parameter space of the model.

\subsection{Vector Mediator}\label{Sec:sChannelGauge}

We now discuss dark matter freeze-in in the simplified model with a vector mediator, $Z'$. We first discuss the case of a $Z'$ that mixes kinetically with the SM hypercharge, followed by the case of a $Z'$ arising from the gauged $L_\mu-L_\tau$ symmetry. 
The results in this subsection will be qualitatively similar to those in the previous (scalar) subsection, but with some crucial differences. In particular, in the kinetically mixed scenario, due to the dependence of the mixing angle on the ratio of the $Z,Z'$ masses (see Eq.\,(\ref{eq:ZprimeMix})), we will find that the heavier mediator as well as the interference term play a more important role. In the $L_\mu-L_\tau$ scenario, only the $Z'$ contributes to the dark matter abundance. 

\subsubsection{Contributions to Dark Matter Freeze-In}

In the EFT framework, the relevant interactions are derived from 
the four-fermion dimension-6 operators $\frac{1}{\Lambda_L^2}(\bar f\gamma_\mu P_L f)(\bar X\gamma^\mu X)$ and $\frac{1}{\Lambda_R^2}(\bar f\gamma_\mu P_R f)(\bar X\gamma^\mu X)$, with $P_{L,R}$ the left-handed and right-handed projection operators, respectively, and the coefficients $\Lambda_{L,R}$ given by 
\beq
\frac{1}{\Lambda_{L,R}^2}=\frac{g_{L,R\,}g_X}{m_Z^2}+\frac{g'_{L,R}\,g'_X}{m_Z'^2}\,, 
\eeq
with the relevant couplings defined in and below Eq.\,(\ref{eq:Zpcoups}). The dark matter yield from these operators can be estimated as
\beq
Y^{\rm{EFT}}\sim\frac{30\,N_c}{(1.66) \pi^{7}~{g_*}^S \sqrt{g_*^{\rho}}}M_{Pl} T_{RH}^3 \left(\frac{1}{\Lambda_{L}^4}+\frac{1}{\Lambda_{R}^4}\right)\,.
\label{eq:YZEFT}
\eeq 
This estimate is similar to that from the scalar case (Eq.\,(\ref{Yefts})), except for different prefactors due to a different Lorentz structure of the operators. 

The full matrix element for this annihilation process can be written as
\beq
|\mathcal M_{2\to 2}|^2=|\mathcal M_{Z}|^2+|\mathcal M_{Z^\prime}|^2+(\mathcal M_{Z}\mathcal M^*_{Z^\prime}+{\rm{h.c.}})\,,
\eeq
where 
\bea
\label{eq:ZZ}
|{\mathcal M}_{Z}|^2 &=& \frac{(q_Dg_D)^2\sin^2\alpha}{4[(s-m_Z^2)^2+(m_Z\Gamma_Z)^2]} \Bigg( (g_L^2+g_R^2)\Big[16 m_X^2m_f^2(\cos^2\theta-\sin^2\theta) \\ \nonumber
&+&8\,s(m_X^2\sin^2\theta-m_f^2\cos^2\theta)+2s^2(1+\cos^2\theta) \Big] +g_L g_R (32m_X^2m_f^2 + 16 m_f^2 s) \Bigg), 
\eea
\be 
\label{eq:ZpZp}
|{\cal M}_{Z'}|^2 = |{\cal M}_{Z}|^2  ~{\rm with} ~\sin\alpha\to\cos\alpha, \, (m_Z,\Gamma_Z)\to(m_{Z'},\Gamma_{Z'}), \,g_{L,R}\to g^\prime_{L,R}\,, 
\ee
\bea 
\label{eq:ZZpinterference}
&&{\cal M}_{Z}{\cal M}^*_{Z'} + {\rm h.c.}= \frac{A~(q_Dg_D)^2\sin\alpha\cos\alpha}{2[A^2+B^2]}\Bigg( (g_L g'_L+g_R g'_R) \times \\\nonumber
&&\Big[16 m_X^2m_f^2(\cos^2\theta-\sin^2\theta) +8\,s(m_X^2\sin^2\theta-m_f^2\cos^2\theta)+2s^2(1+\cos^2\theta) \Big] \\ \nonumber
&&+ (g_L g'_R + g_R g'_L) (16m_X^2m_f^2 + 8 m_f^2 s)\Bigg)\,,
\eea
with
\bea
\label{eq:AB}
&& A = s^2-s(m_Z^2+m_{Z'}^2) + m_Z^2 m_{Z'}^2 + m_Z m_{Z'}\Gamma_Z\Gamma_{Z'}, \nonumber \\
&& B = s(\Gamma_Z m_Z - \Gamma_{Z'}m_{Z'}) + m_Z^2 m_{Z'}\Gamma_{Z'} - m_{Z'}^2 m_Z \Gamma_Z~. 
\eea
 Since $\sin\alpha\approx\tan\alpha\sim \epsilon\frac{m_Z^2}{m_{ Z^\prime}^2}\sin\theta_W$ for $\epsilon\ll 1$ and $m_{Z^\prime}\gg m_Z$, all of these squared matrix elements scale as $\sim \epsilon^2 m_Z^4/m_{Z'}^4$ and therefore are of comparable importance. Note that this is in contrast to the scalar mediator case, where the mixing angle $\sin\theta_h$ can be set independent of the heavy mediator mass, and we chose to fix it to a constant value (see the discussion below Eq.\,(\ref{eq:matrixelement})).  

\subsubsection{Cosmological History}

\begin{figure}[t]
\begin{center}
   \includegraphics[width=.9\linewidth]{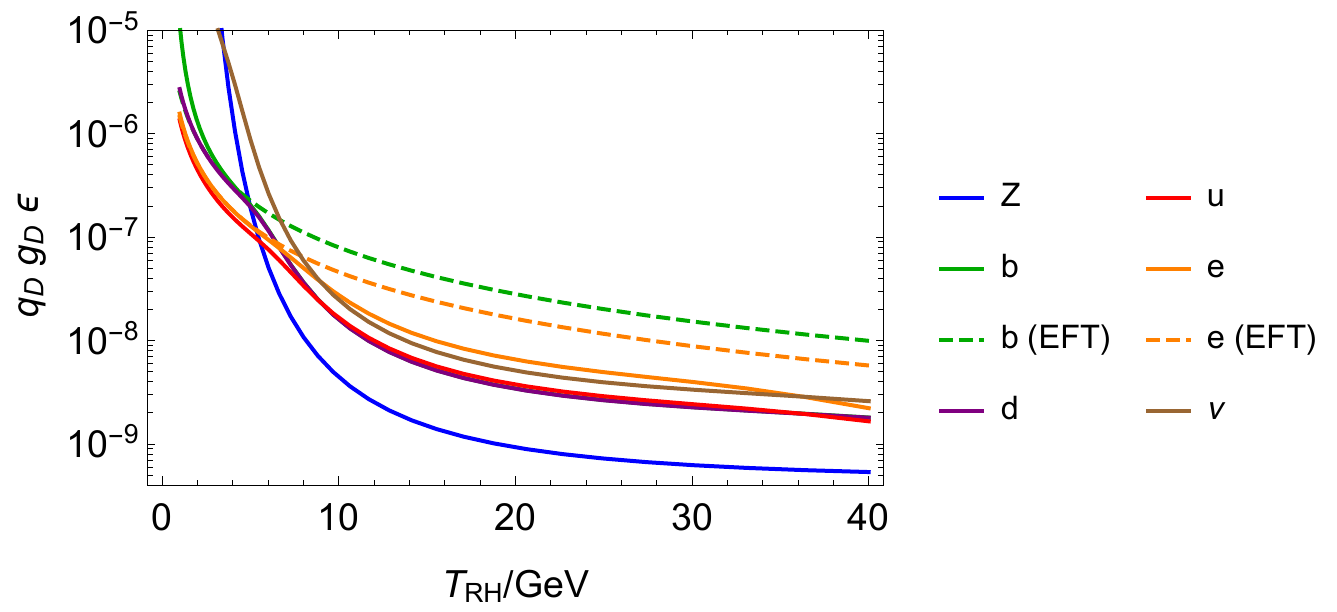}
  \end{center}
  \caption{Relative contributions of various channels for a $Z'$ mediator that kinetically mixes with the SM hypercharge, for $m_{Z^\prime}=500$ GeV, $m_{X}=1$ GeV. We include both the $Z$ and $Z'$ mediated diagrams. We only show a subset of the EFT (non-resonant) contributions, for the bottom quark (dashed green) and the electron (dashed orange). For illustrative purposes, we also include the contribution from the $Z$ decay. The decay contribution from a thermal $Z'$ population does not feature on this plot as it is independent of $\epsilon$ at leading order.}
\label{fig:relativecontributionsZ}
\end{figure}

In Fig.\,\ref{fig:relativecontributionsZ}, we plot the size of the coupling combination $q_D\,g_D\,\epsilon$ needed to produce the observed dark matter relic abundance as a function of $T_{RH}$ for various channels 
for $m_{Z'}=500$ GeV, $m_X=1$ GeV. As in Fig.\,\ref{fig:relativecontributions1d}, the lowest curve represents the dominant production channel. The solid curves represent the result of the full calculation. For illustrative purposes, we show the EFT calculation for $b\bar{b}$ and $e\bar{e}$ annihilations only (dashed curves). At low $T_{RH}\lesssim 7$ GeV, non-resonant annihilation is dominant, with the largest contributions coming from the electron and the up quark \footnote{Since we retain the fermion mass dependence in our calculations,  other fermions become less important at these low temperatures. The neutrino contribution is anomalously weaker due to destructive interference effects at small $\sqrt{s}$.}.  At higher $T_{RH}\gtrsim 7$ GeV, resonant effects start to become important, leading to departures from the EFT approximation. At these temperatures, the behavior is instead matched by the decay of a thermal $Z$ population (blue curve obtained from Eq.\,(\ref{eq:HDecaystoDM})). Overall, couplings $\sim\mathcal{O} (10^{-7})$  are needed to produce the correct relic abundance from non-resonant annihilation at low $T_{RH}\lesssim 7$ GeV, whereas $\sim\mathcal{O} (10^{-9})$ couplings are sufficient for resonant annihilation at $T_{RH}\gtrsim 10$ GeV.

\begin{figure}[t]
\begin{center}
     \includegraphics[width=.85\linewidth]{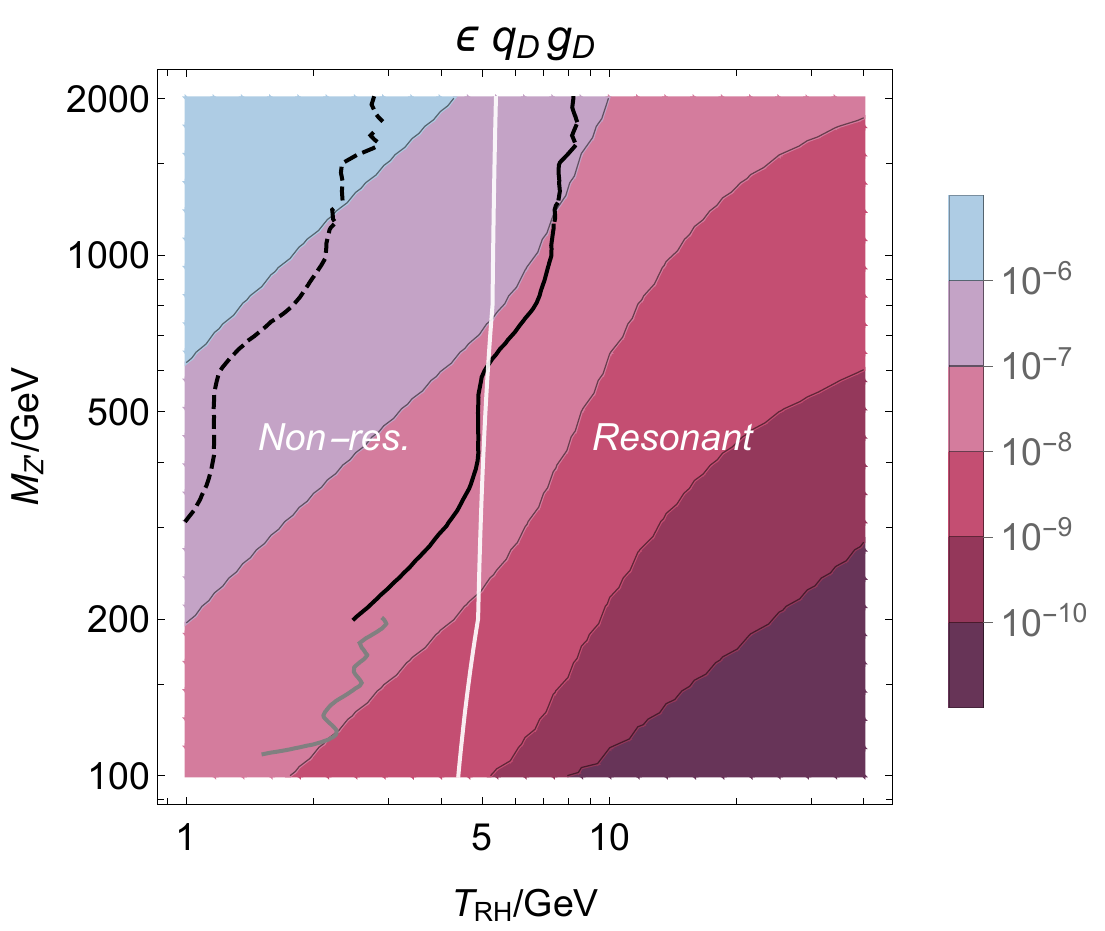}
  \end{center}
	\caption{Value of $\epsilon g_D q_D$ needed to obtain the measured relic abundance as a function of $m_{Z'}$ and $T_{RH}$ for $m_{X}=1$ GeV. The white curve separates regions of parameter space where different contributions dominate dark matter production, as specified by the labels. We also show bounds from LHC searches for a heavy $Z'$ decaying into a lepton pair: the grey curve represents the CMS bound \cite{Sirunyan:2019wqq}; the black curves represent the stronger bound between the ATLAS  \cite{Aad:2019fac} and CMS \cite{CMS:2019tbu} searches. 
	For the LHC bounds, we fix $g_D q_D=3\times10^{-6}$ (solid curves) and $3\times10^{-5}$ (dashed curve). } 
\label{fig:contoursZ}
\end{figure}

In Fig.\,\ref{fig:contoursZ}, we show contours of the values of $\epsilon g_D q_D$ needed to obtain the measured relic abundance as a function of $T_{RH}$ and the heavy mediator mass $m_{Z^\prime}$ for $m_{X}=1$ GeV. A large range of couplings, $\mathcal{O}(10^{-11}-10^{-5}$) can give the correct dark matter abundance. We also show the boundary between regions where non-resonant annihilation dominates, so that the EFT approach gives a good approximation of the yield, and where resonant effects become dominant, and the full calculation must be performed. This boundary occurs at $T_{RH}\sim 5$ GeV and is essentially insensitive to the exact value of $m_{Z'}$. 
We also show bounds from LHC searches for a heavy $Z'$ (see caption of the figure and Sec.\,\ref{sec:colliderpheno} for details) for two different sets of parameters, $g_D q_D=3\times10^{-6}$ (solid curves) and $3\times10^{-5}$ (dashed curve). Thus LHC constraints can be quite strong in the region of parameter space of interest if $g_D q_D$ is small. 


\subsubsection{Modified Vector Mediator: $L_{\mu}-L_{\tau}$}
In this section, we study dark matter freeze-in in the $L_{\mu}-L_{\tau}$ model. The main difference between this framework and the kinetically mixed $Z^\prime$ model is that the heavy $Z^\prime$ mediator here couples directly to both dark matter and SM particles (muon and tau leptons and neutrinos) without requiring mixing with the SM $Z$ boson. We have checked that processes mediated by the SM $Z$ boson, which acquires a coupling to DM via loop-suppressed mixing effects (see Eq.\,(\ref{eq:mutaukinetic})), are suppressed and negligible. Therefore, $Z'$ mediated processes dominate the dark matter production and phenomenology. 

\begin{figure}[t]
\begin{center}
  \includegraphics[width=.65\linewidth]{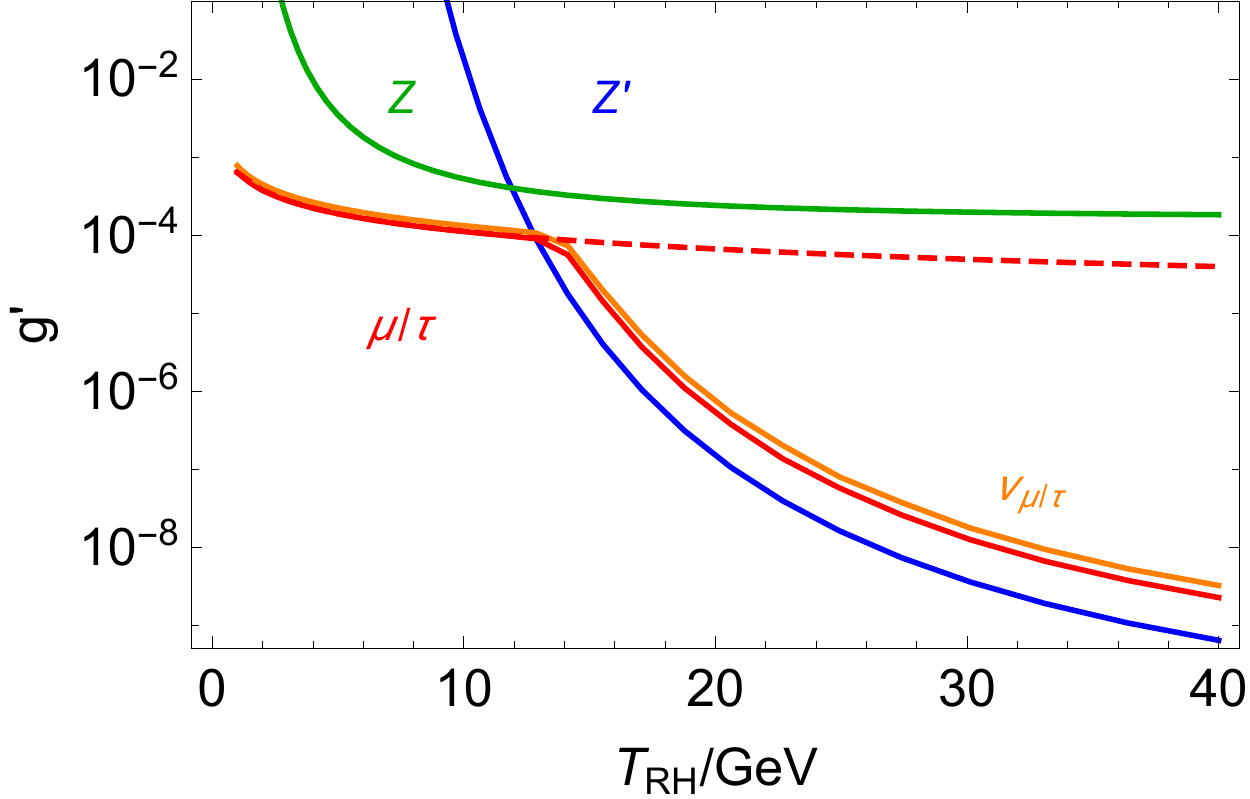}
\end{center}
\caption{ Value of the gauge coupling $g'$ needed to achieve the correct relic abundance of $m_{X}=1$ GeV dark matter from individual channels in the $L_{\mu}-L_{\tau}$ model for $q_X=q_l=1$ and $m_{Z'}=500$ GeV. Solid curves represent the full calculation, the red dashed curve is the EFT (non-resonant) result for the muons and taus, and the blue curve denotes the contribution from $Z'$ decays. The $Z$ decay contribution (green curve) is suppressed due to kinetic mixing only being induced at the loop-level. }
\label{fig:gprime}
\end{figure}

The various contributions to the DM abundance are shown in Fig.\,\ref{fig:gprime}, where we set $q_X=q_l=1$ for simplicity. The solid curves denote the full calculation, while the dashed curve denotes the EFT (non-resonant) treatment (shown only for the muons and taus). The solid blue and green curves represent the contributions from the decays of thermal populations of $Z'$ and $Z$ bosons, respectively. We see that contributions from $Z$ decays are always subdominant due to the loop suppression of the kinetic mixing that gives rise to such decays. Otherwise, in line with previous results, non-resonant contributions dominate at low $T_{RH}\lesssim 15$ GeV, where the correct abundance is obtained for $g'\sim 10^{-4}$. For higher $T_{RH}$, the resonant behaviour is important, and the result is instead captured closely by considering decays of a thermally suppressed abundance of $Z'$ bosons, which gives the correct relic abundance for much smaller couplings. 
The couplings involved in producing the correct relic abundance in this model are larger than those involved in the kinetic mixing case due to the smaller couplings of the mediator with the SM particles as well as the absence of the $Z$-mediated interactions.

\subsection{Salient Features}

We now discuss some salient features common to all s-channel mediator frameworks. 

\begin{figure}[t]
\begin{center}
 \includegraphics[width=.65\linewidth]{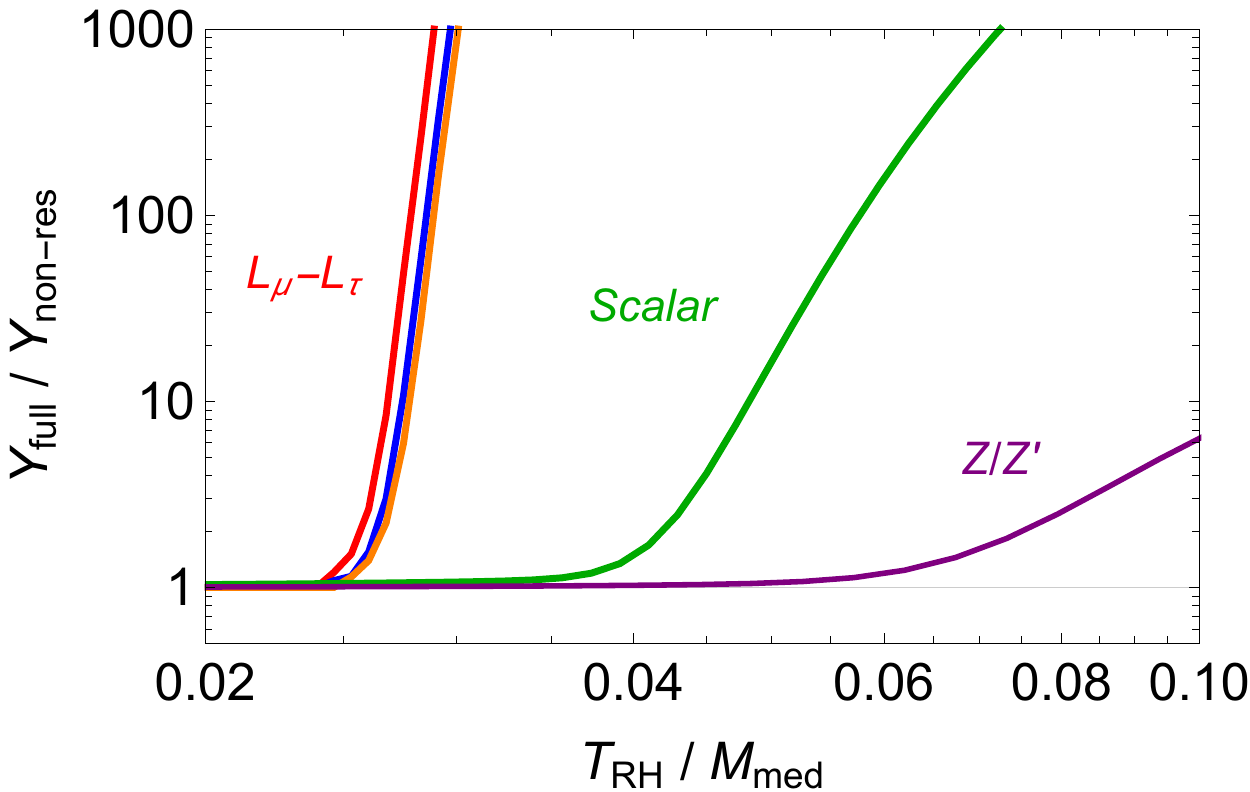}
\end{center}
	\caption{Ratio of the dark matter relic abundance from the full calculation to that obtained in the non-resonant EFT approximation, as a function of $T_{RH}/M_{med}$. From left to right, the curves are for the $L_{\mu}-L_{\tau}$ model with $M_{med}=m_{Z'}=0.1,1,10$ TeV (red, blue, orange curves respectively), the scalar model (green curve, with $M_{med}=m_h$) and $Z'$ kinetic mixing model  (purple curve, with $M_{med}=m_Z$). For all curves, the couplings are chosen such that the full calculation gives the correct relic abundance for $m_X=1$ GeV. }
\label{fig:ratios}
\end{figure}

As we saw in the previous subsections, the main feature is the s-channel resonance, which can dominate the dark matter production process and invalidate the EFT approach. The importance of this effect depends on the couplings involved as well as on the nature of the mediator (whether it couples to both dark matter and the SM directly or via mixing with some SM particle). In Fig.\,\ref{fig:ratios}, we show this feature for the various mediators we have discussed. We plot the ratio $Y_{full}/Y_{non-res}$, where $Y_{full}$ is the dark matter relic abundance from the full calculation (including resonance effects), whereas $Y_{non-res}$ is the abundance obtained from the EFT approximations (obtained by dropping the $s$ dependence in the denominators of the matrix elements). We use the values of the couplings that give the correct relic density for $m_X=1$ GeV with the full calculation. This ratio is plotted as a function of $T_{RH}/M_{med}$, where $M_{med}$ is the mass of the mediator that provides the largest contribution. From left to right, the curves are for the $L_{\mu}-L_{\tau}$ model with $M_{med}=m_{Z'}=0.1,1,10$ TeV (red, blue, orange curves respectively), the scalar model (green curve, with $M_{med}=m_h$) and $Z'$ kinetic mixing model  (purple curve, with $M_{med}=m_Z$). 

At low values of $T_{RH}/M_{med}$, $Y_{full}/Y_{non-res}\approx1$ for all curves, showing that the EFT gives the correct dark matter relic abundance for sufficiently low $T_{RH}$ in all cases.  As $T_{RH}/M_{med}$ increases, an increasingly larger fraction of the thermal distribution of SM fermions can access the s-channel resonance regime, resulting in deviations from the EFT calculations. Recall that the matrix elements scale as $|{\mathcal{M}}|^2 \sim 1/M_{med}^4$ in the non-resonant EFT limit but as $|{\mathcal{M}}|^2 \sim 1/M_{med}^2 \Gamma_{med}^2$ in the resonant regime. Therefore, the deviation from the EFT result is controlled by the magnitude of  $\Gamma_{med}$ relative to $M_{med}$: smaller widths, \textit{i.e.} smaller values of $\Gamma_{med}/M_{med}$ result in earlier deviations from the $Y_{full}/Y_{non-res}\approx1$ limit. This is indeed visible in the plot: the $Z'$ boson in the $L_\mu-L_\tau$ model has tiny couplings to the dark and SM particles, therefore a very narrow width, and thus begins to deviate from the EFT calculation already at $T_{RH}/M_{Z'}\approx 0.025$. The scalar curve deviates next at $T_{RH}/M_{h}\approx 0.035$, since the SM Higgs also has a relatively narrow width. Finally, the $Z/Z'$ curve deviates from the EFT result at $T_{RH}/M_{Z}\approx 0.05$, since the $Z$ width is larger. This implies that for dark matter freeze-in in s-channel models, \textit{the EFT approach already breaks down when the mediator mass is one or two orders of magnitude above the reheat temperature}. The relative steepness of the curves is also determined by the size of the width relative to the mass of the mediator: the smaller the width, the faster the rate of departure from the EFT result. Finally, for the $L_\mu-L_\tau$ model, we have shown three curves, for $m_{Z'}=0.1,1,10$ TeV (red, blue, orange curves respectively). Consistent with the discussion above, a lighter ${Z'}$ departs from the EFT result at  lower $T_{RH}/M_{Z^\prime}$ since it requires smaller couplings to obtain the correct relic density, hence $\Gamma_{med}/M_{med}$ is smaller. 

\begin{figure}[t]
\begin{center}
   \includegraphics[width=0.65\linewidth]{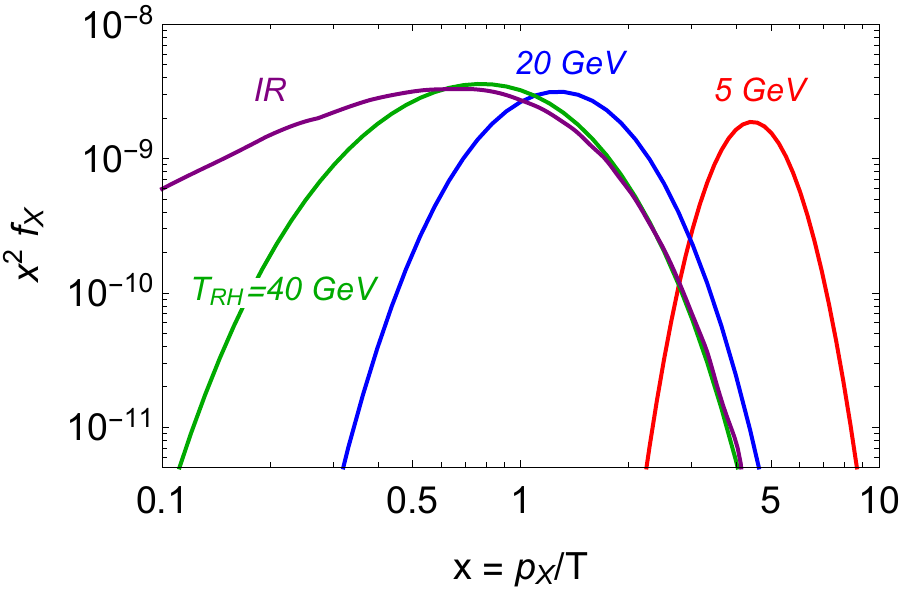}
  \end{center}
  \caption{Dark matter phase space distribution as a function of the DM momentum ($x = p_X/T$) for a DM candidate with mass $m_X=1$ GeV produced from SM Higgs decays, for $T_{RH}=5, 20, 40$ GeV. The normalization of each distribution is set by choosing the coupling $y_{hXX}$ that gives the correct relic abundance at the various $T_{RH}$ values. The curve labelled ``IR" (adapted from Ref.\,\cite{Roland:2016gli}) represents the standard IR freeze-in scenario with $T_{RH}\gg M_{med}$, where most of the DM population is produced at $T\approx M_{med}$.
}
\label{fig:decayphase}
\end{figure}

The prominence of the s-channel resonance with heavy mediators significantly changes not only the relic abundance of dark matter but also its momentum distribution. This is illustrated in Fig.\,\ref{fig:decayphase}, where we plot the DM momentum distribution from SM Higgs decays (representative of resonant annihilation)\footnote{Computing the distributions from the full fermion annihilation is numerically challenging, but we note that they are expected to follow the same shapes as the decay curves when the resonant behavior is important, but with some smearing due to the broader thermal distributions of SM fermions.}  for various values of $T_{RH}=5, 20, 40$ GeV.  In general, dark matter momentum distributions peak at $p/T\!\sim\!1$ since dark matter particles are produced, on average, with energy corresponding to the temperature of the thermal bath. However, when the process is dominated by resonant annihilation via the s-channel Higgs mediator, dark matter particles get produced dominantly at $p\sim m_h/2$. Since the temperature of the thermal bath is much lower, the DM momentum distribution peaks in this case at $p/T\sim m_h/(2 T_{RH})$. This ratio gets further suppressed by an $\mathcal{O}(1)$ factor as entropy released from subsequent decoupling and decays of the SM particles heats the thermal bath but not the DM population. The various curves in the figure agree with these considerations. For comparison, we also show the momentum distribution expected in standard IR freeze-in scenarios (purple curve, adapted from Ref.\,\cite{Roland:2016gli}), where $T_{RH}\gg M_{med}$. In this scenario, most of the dark matter production occurs at $T\gtrsim M_{med}$, and consequently the IR freeze-in distribution is colder than the other distributions with smaller $T_{RH}$, as well as broader with a larger lower momentum counterpart. Such modified momenta distributions provide another observable difference between standard IR freeze-in scenarios and freeze-in with a heavy mediator and a low reheat temperature.

\section{Dark Matter Freeze-In: T-Channel Mediator}
\label{sec:tchannel}

We now focus on dark matter freeze-in with a t-channel mediator. This scenario contains several qualitative differences from the s-channel mediator framework discussed in the previous section. In particular, the t-channel framework does not feature the resonant behavior of the annihilation processes that resulted in large deviations from the EFT approach in the s-channel scenario. Instead, as we will see, deviations from the EFT arise due to on-shell production of the t-channel mediator through annihilation processes, as well as new loop-induced decays of SM particles. As mentioned earlier, we study the t-channel scenario in two cases: (1) mediator coupling to the right-handed electron, $e_R$; (2) mediator coupling to the right-handed top quark, $t_R$. 

\begin{figure}[t!]
\begin{center}
  \includegraphics[width=1\linewidth]{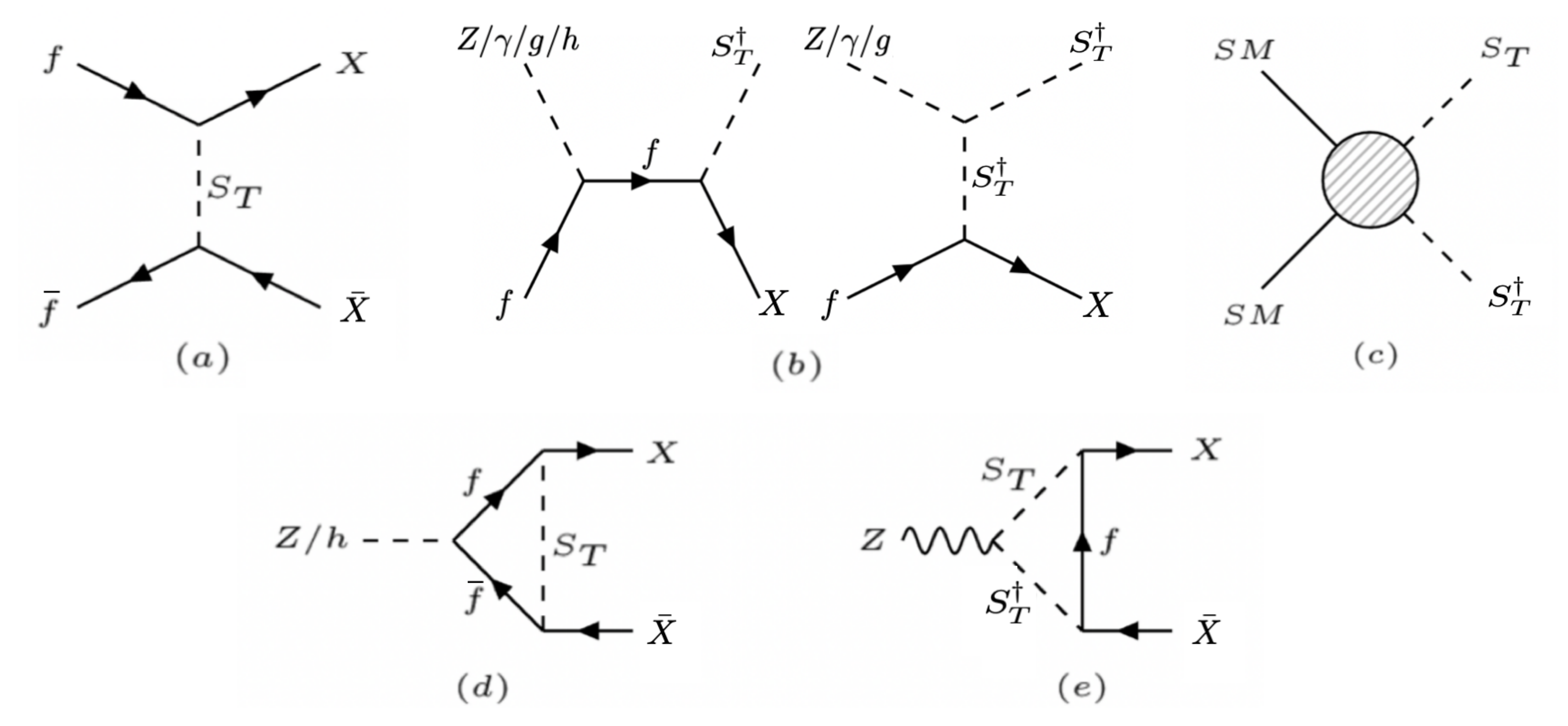}
  \end{center}
\caption{Various contributions to dark matter freeze-in in the t-channel mediator scenario. Diagram (c) is representative of several processes involving SM particles annihilating into a pair of mediators.}
\label{fig:Tdiagrams}
\end{figure}

\subsection{Contributions to Dark Matter Freeze-In}

The various processes contributing to dark matter production in the t-channel framework are shown schematically in Fig.\,\ref{fig:Tdiagrams}. In the low energy EFT, DM freeze-in arises via four-fermion interactions obtained by integrating out the heavy mediator $S_T$ (diagram (a)\,), analogous to the s-channel case. However, since DM is now no longer $Z_2$-symmetric, and since $S_T$ carries SM charges, additional contributions exist and must be taken into account. Since the $S_T-X$ system carries an effective $Z_2$-symmetry, there are three classes of annihilation processes: $X\bar{X},\,X\,S_T^\dagger,$ or $S_T\,S_T^\dagger$ (top row in Fig.\,\ref{fig:Tdiagrams}). Since $S_T\to \bar{f} X$ is the only decay channel available for $S_T$ at tree-level, each $S_T$ particle produced in the early Universe will result in a DM particle, hence the latter two diagrams contribute to secondary DM production. In addition to these annihilation processes, the mediator $S_T$ also gives rise to novel loop-level processes that produce DM via decays of SM particles. This is shown in the second row of Fig.\,\ref{fig:Tdiagrams}. We now discuss these various contributions in more detail. 

\vskip 2mm 
\noindent\textbf{(a) $f\bar{f}\to X\bar X$ annihilation}

As in the s-channel case, we can estimate the DM yield from freeze-in via a four-fermion dimension-6 operator $\frac{1}{\Lambda^2} (f\bar{X})(X \bar f)$ obtained from integrating out the t-channel mediator from the Lagrangian in Eq.\,(\ref{eq:lagrangiantchannel}), where now $\Lambda=m_{S_T}/y_T$
\beq
Y^{EFT}\sim\frac{15\,N_c}{(1.66) \pi^{7}~{g_*}^S \sqrt{g_*^{\rho}}}\frac{y_T^4 \,M_{Pl} T_{RH}^3}{m_{S_T}^4}\,
\label{eq:tchanneleft}
\eeq
assuming $m_f\ll T_{RH}$. For the full calculation, we use the full amplitude for the process,
\begin{align}
	|\mathcal{M}|^2 = \frac{y_T^4 ((m_f + m_X)^2-t)^2}{(m_{S_T}^2 - t)^2},
\end{align}
where $t$ is the standard Mandelstam variable. Note that there is no resonance effect for a t-channel mediator, as was the case for the s-channel mediator, as the propagator cannot go on-shell. Thus, Eq.\,(\ref{eq:tchanneleft}) is expected to remain a good approximation of the annihilation contribution.

\vskip 2mm 
\noindent\textbf{(b) $(Z/\gamma/g/h)\,f \to S_T^\dagger X$ coannihilation}

Fermion coannihliation with an electrically neutral SM boson $(Z/\gamma/g/h) f \to S_T^\dagger  X$ proceeds via the two diagrams shown in Fig.\,\ref{fig:Tdiagrams}\,(b). In contrast to the $f\bar f\to X\bar X$ process, the cross section for this class of diagrams scales as $y_T^2$ instead of $y_T^4$. However, since these processes involve the mediator being produced on-shell and we have assumed $T_{RH}<m_{S_T}$, the probability for such interactions to occur is Boltzmann suppressed.  Note that all initial state particles, including $Z, h, t$, can be approximated to be massless since their masses are negligible compared to the kinematic threshold required for this process. 

These interactions, if sufficiently rapid, will produce a thermal abundance of dark matter. Hence we require that these interactions remain slower than Hubble at all times, $ \sum n \langle\sigma v\rangle \!<\! H$, where the sum is over all processes that contribute to the production of the $X S_T^\dagger$ final state. For the case where $S_T$ couples to $t_R$, the dominant process is coannihilation with a gluon, $g\,t_R \to S_T^\dagger X$ (the $h \,t_L\to S_T^\dagger X$ contribution is of the same order of magnitude due to the large top Yukawa), 
 which enforces the approximate condition
 \beq
 \alpha_s\,y_T^2\,\sqrt{x} e^{-x}\lesssim m_{S_T} /\,M_{Pl},
\label{eq:tthermalization}
\eeq 
where $x=m_{S_T}/T_{RH}$, and we have ignored $\mathcal{O}(1)$ factors in the estimate. For the case where $S_T$ couples to $e_R$, coannihilations with $Z/\gamma$ are the most important: summing these contributions, the approximate condition to avoid thermalization is the same as in Eq.\,(\ref{eq:tthermalization}) with $\alpha_s\to\alpha$.

We focus on regions of parameter space where the non-thermalization conditions are satisfied, so that dark matter is produced from freeze-in.  In this case, freeze-in formulae analogous to Eq.\,(\ref{eq:abundanceGeneral}) (see \textit{e.g.}~discussions in \cite{Elahi:2014fsa}) can be used to calculate the freeze-in abundance. 
For instance, for the process $h\,t_L \to S_T^\dagger X$, the first diagram in Fig.\,\ref{fig:Tdiagrams}\,(b) gives
\beq\label{eq:YVt1}
Y_{X\,(h\, t_L \to S_T^\dagger  X)}\approx \frac{45~ g_T^2~ y_T^2\,M_{Pl}}{(1.66) 2^{11} \pi^7 ~ {g_*}^S \sqrt{g_*^{\rho}}} \int^{T_{RH}}_0 \frac{dT}{T^5} \int^\infty_{m^2_{S_T}}ds ~ \frac{(s-m^2_{S_T})^{2}}{s^{3/2}} \, K_1(\sqrt{s}/T).
\eeq
For this process, $g_T^2=3 (m_t/v)^2$ \footnote{In the case of the Higgs we do not include the contribution from the second diagram in Fig.\,\ref{fig:Tdiagrams}\,(b), as it depends on the coupling $hS_TS_T^\dagger$, which is an independent parameter of the Lagrangian. We neglect this coupling, for simplicity. However, in general, this diagram will contribute to the dark matter abundance, especially at relatively sizable values of $T_{RH}$. However, even in this regime, we do not expect a sizable change in the value of $y_T$ needed to achieve the correct relic abundance compared to what is shown in Fig.\,\ref{fig:Tmediator}.}. For the other processes, both diagrams in Fig.\,\ref{fig:Tdiagrams}\,(b) contribute, and the DM yield does not admit a simple closed form as above, but they are straightforward to evaluate numerically. 

\vskip 2mm 
\noindent\textbf{(c) $SM \to S_T S_T^\dagger$}

The final diagram in the top row of Fig.\,\ref{fig:Tdiagrams} represents annihilations of SM particles that pair-produce the mediator. There are several classes of diagrams contributing with the most important being annihilations of gauge bosons. Such dimension-4 operators involve no small couplings and can therefore be efficient enough to thermalize the $S_T$ population if $T_{RH}$ is sufficiently high. This occurs if the reheat temperature is greater than the $S_T$ freeze-out temperature $T_{f.o.}$, which can be very roughly estimated as $T_{f.o.}\approx m_{S_T}/20$ for weak scale interactions. In this regime, the dark matter abundance is a result of thermal freeze-out of the mediator, whose decays then populate dark matter, rather than freeze-in processes. 

For simplicity, we will focus on the $T_{RH}<T_{f.o.}\approx m_{S_T}/20$ regime, and assume that $S_T$ is never in equilibrium with the SM bath\,\footnote{If $S_T$ interacts with gluons, as in the case where it couples to $t_R$,  thermal equilibrium is maintained to far lower temperatures. However, in this case the $S_T$ freeze-out abundance is also significantly smaller, and the subsequent contribution to dark matter abundance becomes negligible.} but instead can be produced via freeze-in. Note that the DM abundance from this contribution is independent of $y_T$. However, given that the $S_T$ abundance from freeze-in has to be smaller than its thermal freeze-out abundance, such freeze-in contributions are very small and generally do not contribute significantly to the dark matter relic abundance.

\vskip 2mm 
\noindent\textbf{(d,e) $Z,h$ decay}

The SM Higgs and $Z$ bosons can decay into a pair of dark matter particles through one-loop diagrams with $f, S_T$ in the loop, as shown in the bottom panel of Fig.\,\ref{fig:Tdiagrams}. We compute the decay widths with the help of Package-X \cite{Patel:2015tea,Patel:2016fam}. 

In the model in which $S_T$ couples to $e_R$, the $Z\to X\bar{X}$ decay width in the $m_e,\,m_Z\ll m_{S_T}$ limit can be written as
\begin{equation}
	\Gamma_{Z\to X\bar X} \simeq \frac{y^4_T g_T^2}{8 \pi m^2_Z} ~\sqrt{\frac{m^2_Z}{4}-m^2_X} ~\frac{m^6_Z}{2^7 3^3 \pi^2 m^4_{S_T}} \left[1 + \frac{1}{3 \pi^2}\text{log}\left(\frac{m^2_{S_T}}{m^2_Z}\right) + \frac{1}{\pi^2}\text{log}\left(\frac{m^2_{S_T}}{m^2_Z}\right)^2\right]\,,
	\label{eq:zloopwidth}
\end{equation}
where $g_T=e\tan\theta_W$ is the $Z$ coupling to right-handed electrons, and we have dropped terms suppressed by powers of $m_e^2/m_{S_T}^2$. In the model in which $S_T$ couples to $t_R$, the corresponding width receives additional comparable contributions that consist of powers of $m_t^2/m_{S_T}^2$ instead of $m_Z^2/m_{S_T}^2$ and depend on the $Z$ coupling to the SM left-handed top quark. We do not present the full expression for this decay width since it is quite lengthy\,\footnote{For $Z$ decays involving top quark loops, the part of the amplitude leading to Eq.\,(\ref{eq:zloopwidth}) is proportional to $g_T \frac{m^2_Z}{m^2_{S_T}}\left[\frac{1}{18} + \frac{1}{3} \text{log}\left(-\frac{m^2_{S_T}}{m^2_Z}\right)\right]$. The additional contributions are proportional to $\frac{m^2_t}{m^2_{S_T}}\left[(g_T^L+g_T)-(g_T^L - g_T) \text{log}\left(-\frac{m^2_Z}{m^2_{S_T}}\right)\right]$, where $g_T^L$ is the SM $Z$ coupling to $t_L$. These terms lead to contributions comparable to the ones in Eq.\,(\ref{eq:zloopwidth}).}, but we use it in our numerical calculations.

We evaluate the Higgs diagram in Fig. \ref{fig:Tdiagrams}\,(d) to be
\begin{equation}
	\Gamma_{h\to X\bar X} = \frac{N_c \, y^4_T}{8 \pi m^2_h} ~\sqrt{\frac{m^2_h}{4}-m^2_X} ~\frac{m^2_X(m^2_h - 4 m^2_X) m_f^4}{2^{9} \pi^4 m^4_{S_T} v^2} \,,
	\label{eq:hloopwidth}
\end{equation}
for both models ($f=e_R,t_R$, with $N_c=1,3$, respectively). 

The dark matter abundance from such decays can be estimated by substituting the above decay widths into Eq.\,(\ref{eq:HDecaystoDM}). Note that while the $Z$ decay width can be enhanced by large logarithms, the Higgs decay width is suppressed by powers of both dark matter and loop fermion masses, as the process requires a helicity flip 
since the mediator couples only to right-handed fermions.\,\footnote{For simplicity, for the Higgs decay, we ignore the possible coupling between the Higgs and $S_T$ of the form $\lambda vh S_TS_T^\dagger$, which would introduce a new diagram analogous to Fig.\,\ref{fig:Tdiagrams}\,(e). This contribution would scale as $\sim (\lambda v^2)^2$ instead of  $(m^2_f)^2$ in Eq.\,(\ref{eq:hloopwidth}), and can be important. However, since the Higgs contribution to the relic abundance turns out to be subdominant in all cases (see Fig. \ref{fig:Tmediator}), this additional contribution would not change our conclusions.} Due to these additional suppression factors, as well as a smaller number density of Higgs bosons compared to $Z$ bosons in the early Universe at $T_{RH}<m_h, m_Z$, the contribution from Higgs decay is several orders of magnitude smaller than the contribution from $Z$ decay in all cases we consider (see Fig.\,\ref{fig:Tmediator}).

\subsection{Cosmological History}

\begin{figure}[t]
\begin{center}
  \includegraphics[width=.85\linewidth]{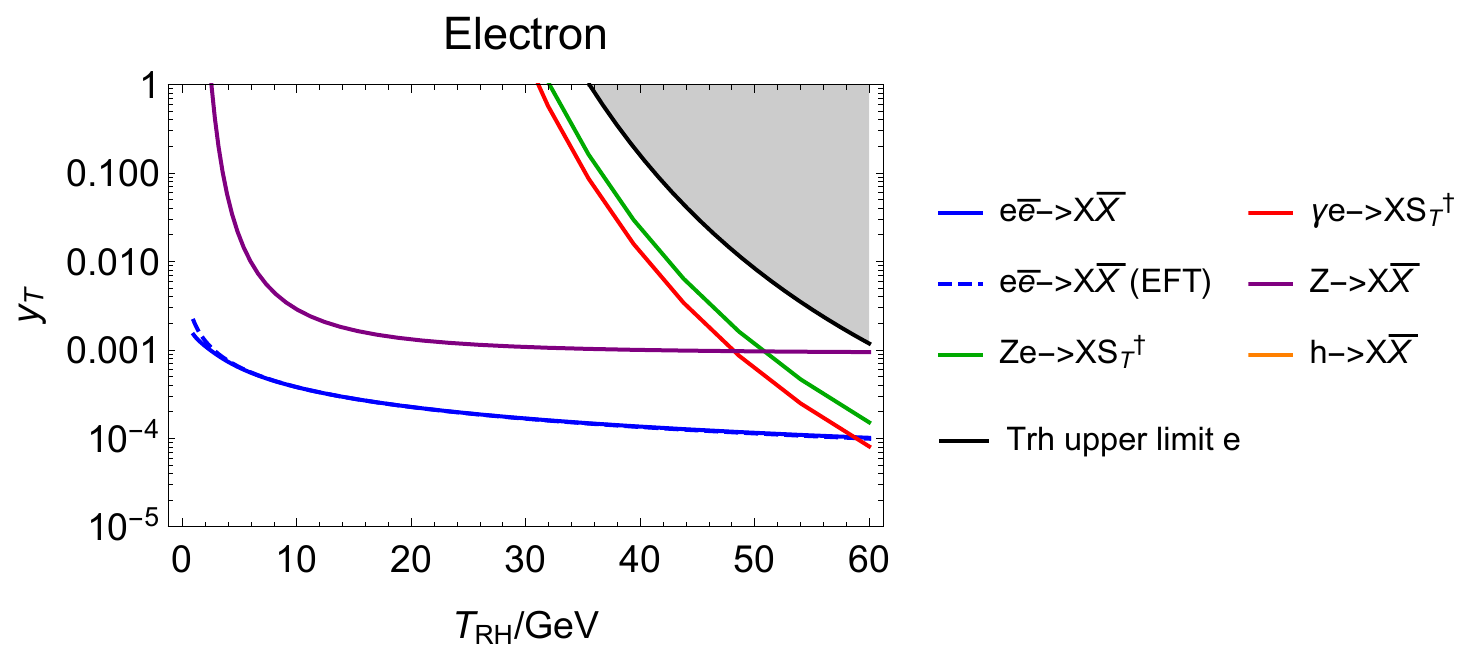}\\
   \includegraphics[width=.85\linewidth]{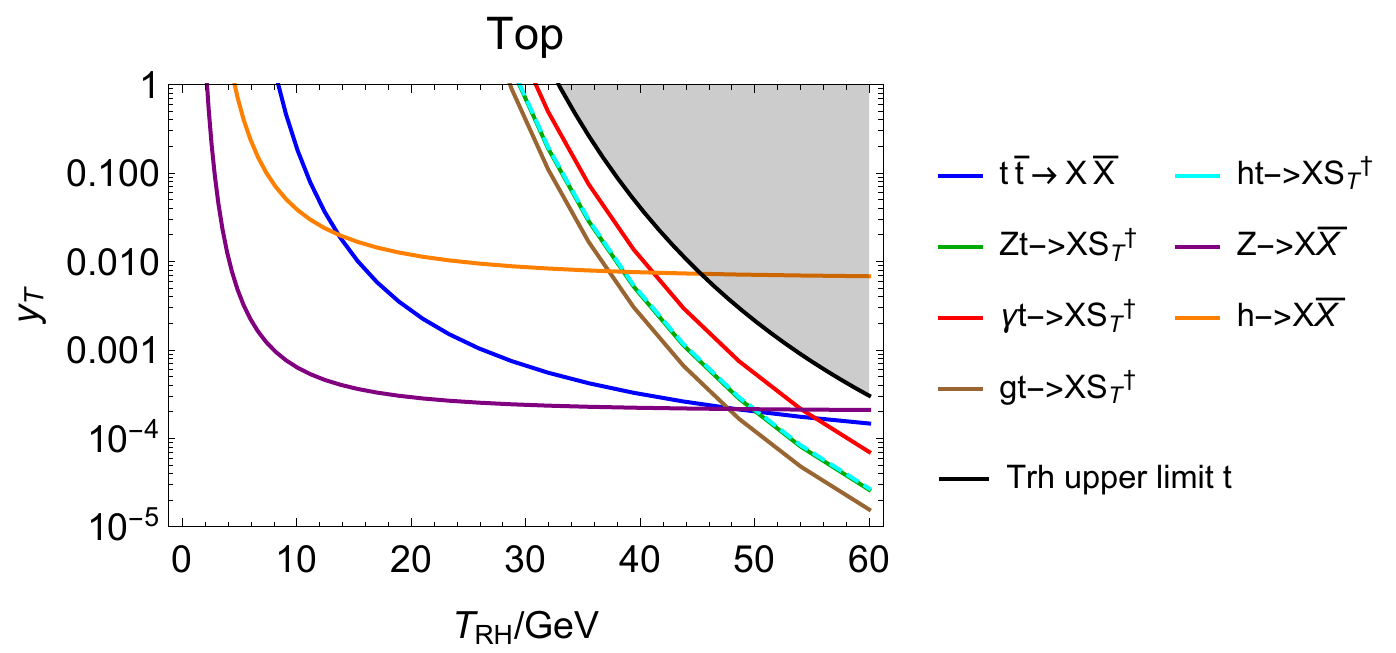}
  \end{center}
  \caption{The size of the coupling $y_T$ needed to produce the correct relic density from various processes as a function of the reheat temperature, $T_{RH}$, for $S_T$ coupling to the right-handed electron $e_R$ (top panel) and to the right-handed top-quark $t_R$ (bottom panel), for  $m_{X}=1$ GeV and $m_{S_T}=1200$ GeV. The dashed curve in the electron plot is the EFT approximation. In the shaded regions, coannihilation processes are rapid enough to produce a thermal DM population.}
\label{fig:Tmediator}
\end{figure}

We now turn to a discussion of the interplay between the above contributions in setting the correct dark matter relic density from freeze-in. Fig.\,\ref{fig:Tmediator} shows the relative contributions of the various channels as a function of $T_{RH}$ for $m_{X}=1$ GeV and $m_{S_T}=1200$ GeV for scenarios where the mediator couples to $e_R$ (top panel) or $t_R$ (bottom panel). Here, we choose $m_{S_T}=1200$ GeV for the mediator mass in view of strong constraints from LHC searches (see Sec.\,\ref{sec:tcollider}). 

For the electron case (top panel of Fig.\,\ref{fig:Tmediator}), the EFT approximation from Eq.\,(\ref{eq:tchanneleft}) (dashed blue curve) closely matches the full calculation for $e^+e^-\to X\bar{X}$ (solid blue curve) throughout, as no strong resonance features are present for a t-channel mediator. The contribution from loop-mediated $Z$ decays (purple curve) is seen to be subdominant to this annihilation contribution throughout. Higgs decay contributions (orange curve) are even weaker and do not feature in this plot. At low $T_{RH}$ ($<60$ GeV), fermion annihilation dominates dark matter production, giving the correct relic density for $y_T\sim\mathcal O(10^{-4})$. At higher $T_{RH}$, coannihilation with a SM boson, in this case $\gamma$ and $Z$, become increasingly important, as an increasingly greater fraction of the thermal population gains enough energy to produce $S_T$ via $(\gamma/Z) e\to S_T^\dagger X$. These coannihilation processes overtake fermion annihilation as the dominant dark matter production channel around $T_{RH}\sim 60$ GeV. In the shaded region, the coannihilation processes thermalize the DM with the SM bath as determined by Eq.\,(\ref{eq:tthermalization}) with $\alpha_s\to\alpha$. As anticipated, the curves for the correct relic density from freeze-in lie away from this region. 

Some of these features change when we consider the scenario where the mediator couples to $t_R$ (bottom panel of Fig.\,\ref{fig:Tmediator}). Here, the fermion annihilation curve (solid blue) is significantly weaker (i.e. requires significantly larger couplings) than for the electron case, as the number density of top quarks in the thermal bath is severely Boltzmann suppressed at such low temperatures.\,\footnote{We do not show the analogous approximate solution from Eq.\,(\ref{eq:tchanneleft}), which was derived assuming the initial fermions are massless, which is inapplicable for the top quark.} For this reason, $Z$ loop decays, although suppressed by several factors, provide the dominant contribution to dark matter freeze-in for $T_{RH}\lesssim 50$ GeV, producing the correct relic density for $y_T\sim 10^{-4}-10^{-3}$. The contribution from Higgs decays, enhanced by the large top Yukawa coupling, is also visible, but remains subdominant to the $Z$ decay contribution. As in the $e_R$ case, coannihilation processes with SM bosons become dominant at larger $T_{RH}$. In this case, coannihilation with a gluon, $g t\to S_T^\dagger X$ (brown curve in the figure), dominates for $T_{RH}\gtrsim 50$ GeV, and significantly smaller couplings $\sim 10^{-5}$ can produce the correct relic abundance.

\begin{figure}[t!]
\begin{center}
  \includegraphics[width=.67\linewidth]{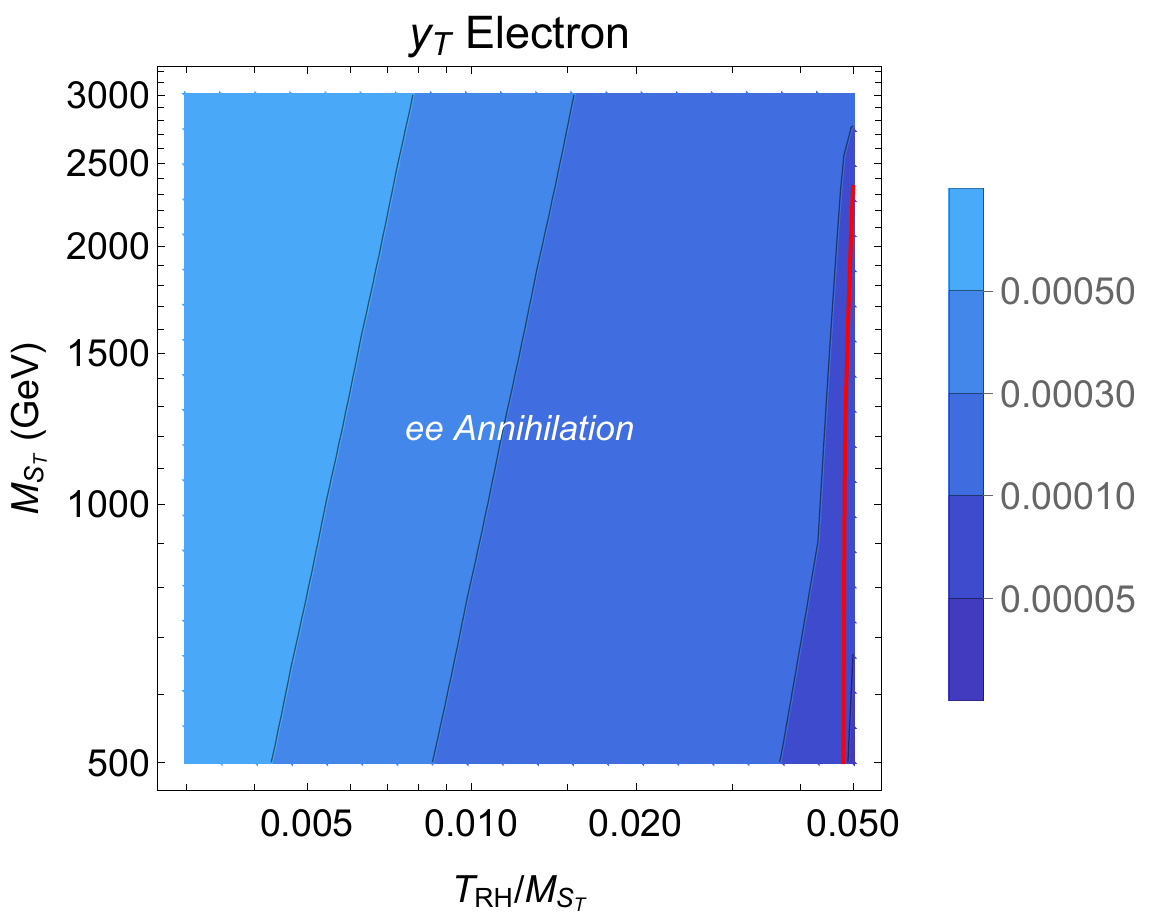}
  \includegraphics[width=0.67\linewidth]{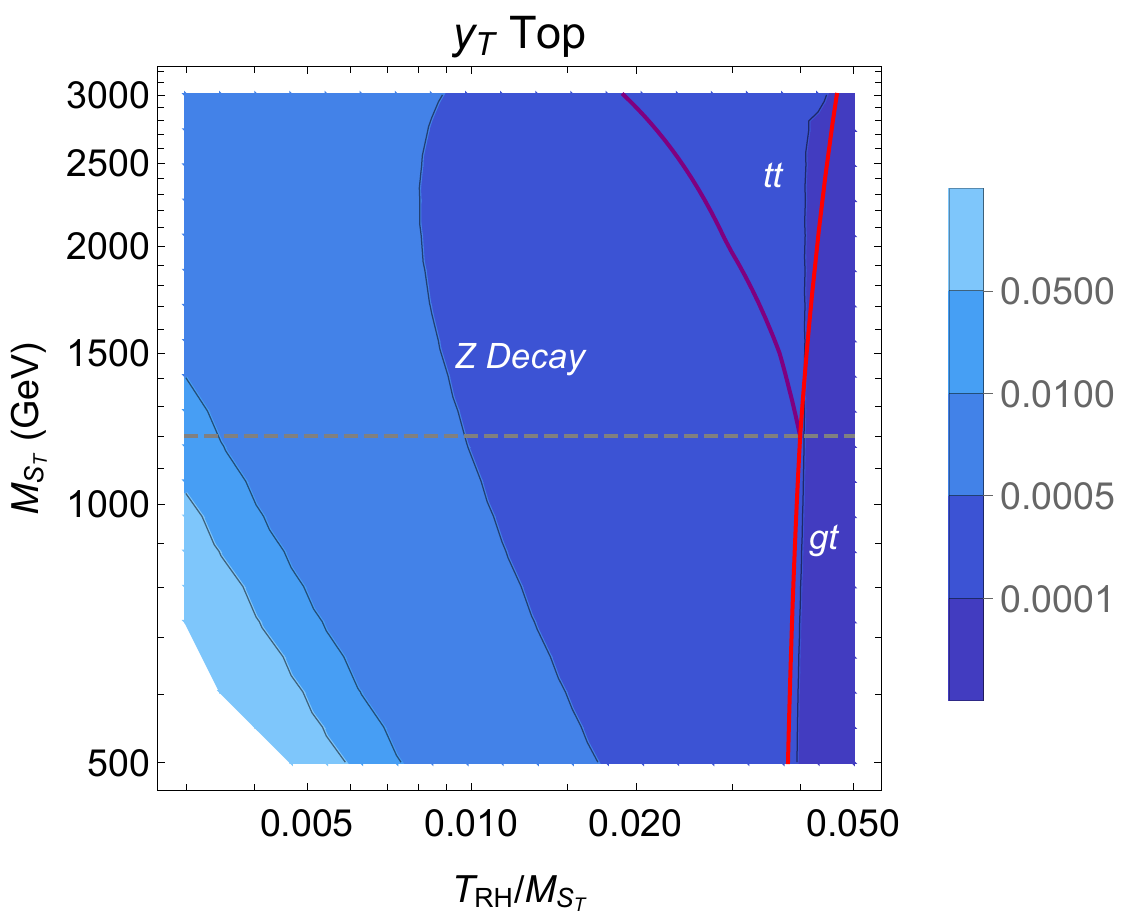}
  \end{center}
\caption{Contours of the coupling $y_T$ needed to produce the correct relic density as a function of the reheat temperature, $T_{RH}$, and mediator mass, $m_{S_T}$, for $m_{X}=1$ GeV, for scenarios where the mediator couples to the right-handed electron (top panel) or the right-handed top quark (bottom panel). Red curves separate regions where different processes dominate dark matter production, as denoted by the labels (see text for more details). The dashed grey curve in the bottom panel denotes the approximate lower bound on $m_{S_T}$ from the LHC.}
\label{fig:Tcomparison}
\end{figure}

In Fig.\,\ref{fig:Tcomparison}, we show contour plots of the value of the coupling $y_T$ needed to obtain the correct relic abundance as a function of the mediator mass $m_{S_T}$ and the reheat temperature $T_{RH}$. As explained earlier, we terminate the x-axis at $T_{RH}/m_{S_T}=1/20$, beyond which the mediator is likely in thermal equilibrium with the SM bath. We also delineate regions of parameter space where different processes dominate dark matter production. For the case of $S_T$ coupling to $e_R$ (top panel), the annihilation process $e^+e^-\to X\bar{X}$ dominates in large regions of parameter space, giving the desired relic abundance with $y_T=\mathcal O( 10^{-4})$. This changes at $T_{RH}/M_{S_T}\approx 0.05$, beyond which the coannihilation process $(\gamma/Z) e\to S_T^\dagger X$ dominates. This boundary is delineated by a red curve in the plot. Thus, the EFT approach where $S_T$ is integrated out breaks down at $T_{RH}/M_{S_T}\approx 0.05$. 

The bottom panel of Fig.\,\ref{fig:Tcomparison} shows the analogous contour plot for $S_T$ coupling to $t_R$. As explained before, in most of the low $T_{RH}$ parameter space, $Z$ decays provide the dominant contribution as $t\bar{t}$ annihilation is Boltzmann suppressed for $T_{RH}\!<\!m_t$. Since the $Z$-decay width into dark matter is parametrically suppressed by the heavy mediator mass as well as loop factors (see Eq.\,(\ref{eq:zloopwidth}) ), somewhat large couplings $\gtrsim 10^{-3}$ are needed to obtain the correct relic density. Indeed, in the white region on the bottom left corner of the plot, the required coupling is $>\!1$. As $T_{RH}$ increases, top-gluon coannihilation, $g t\to S_T^\dagger X$, grows to dominate, and smaller couplings $y_T\sim 10^{-4}$ are sufficient. In the plot, this occurs to the right of the red curve at $T_{RH}/m_{S_T}\approx 0.04$. For $m_{S_T} \gtrsim 1200$ GeV, we also see the emergence of a third region, 
where $t\bar{t}\to X\bar{X}$ dominates: in this region, $T_{RH}$ is sufficiently lower than $m_{S_T}$ that the coannihilation processes involving on-shell production of $S_T$ are suppressed, but high enough that the Boltzmann suppression of the thermal abundance of top quarks is no longer too severe, so that dark matter production from $t\bar{t}$ annihilations are dominant. 
Hence all three processes -- fermion annihilation, fermion-boson coannihilation, and boson decays -- can be the leading dark matter production mode in this t-channel scenario. 

It is instructive to compare the nature of EFT breakdown in the t-channel scenario with those from the s-channel framework (Fig.\,\ref{fig:ratios}). In both cases, the EFT breaks down due to the emergence of processes where the mediator is produced on-shell. In the s-channel scenarios, the mediator is produced on resonance via the inverse decay process $f\bar{f}\to h, S,Z,Z^\prime$, whereas in the t-channel case it is a product of coannihilation between a SM fermion and a boson. 

We end with a brief comment regarding the dark matter momentum distribution in the t-channel mediator scenario. The various contributing processes have distinct energy scales associated with them, and will therefore produce dark matter with different momenta. For the electron case, fermion annihilations produce DM with $p\sim T$, whereas coannihilations produce DM with $p\sim 0$ directly, as well as with $p\sim m_{S_T}/2$ from the subsequent decays of $S_T$. Likewise, the top quark case features annihilations ($p\sim m_t$), coannihilations ($p\sim 0,\,m_{S_T}/2$), as well as $Z$ decays ($p\sim m_Z/2$). Thus, the dark matter momentum distribution carries imprints of the dominant production process. We leave a detailed investigation of such features for future study.

\section{Phenomenology}
\label{sec:pheno}

Having discussed the early Universe history, we now turn to a discussion of the phenomenological aspects of the various frameworks discussed in this paper. It is well known that indirect detection of dark matter annihilations is extremely unlikely in freeze-in models due to the small \textit{effective} couplings involved, which remains true in the setups we studied in this paper. However, while the effective couplings are small, the \textit{real} couplings involved in SM-dark matter interactions, relevant for direct experimental probes such as direct detection and colliders, can be relatively sizable for $T_{RH}\!\ll\! M_{med}$, improving detection prospects on these fronts. We discuss collider prospects below, and follow with a short discussion of direct detection. 

\subsection{\boldmath Probing mediators at collider experiments}
\label{sec:colliderpheno}

Collider phenomenology of standard freeze-in setups often involve displaced decays of the mediator particles due to the feeble couplings involved, see, \textit{e.g.}~\cite{Co:2015pka,DEramo:2017ecx,No:2019gvl,Belanger:2018sti,Garny:2018ali}. For the setups we have considered, which can involve larger couplings, the signatures can be qualitatively different. 

\subsubsection{S-channel scalar mediator}\label{Sec:LHCScalar}

The DM-Higgs coupling, $y_{hXX}$, needed for DM freeze-in will induce an exotic decay channel of the SM Higgs into dark matter with the decay rate
\begin{equation}
\Gamma(h\to X\bar X)=\frac{m_h}{8\pi}y_{hXX}^2\left(1-\frac{4m_{X}^2}{m_h^2}\right)^{3/2}\,.
\end{equation}
The combination of the most recent ATLAS searches for invisible Higgs decays, performed with $5,~20, ~139$ fb$^{-1}$ of 7, 8, and 13 TeV data, sets a bound of BR$(h\to{\rm{invisible}})\sim 0.11$ at 95$\%$ C.L. \cite{ATLAS:2020kdi}. Similarly, the latest CMS combination of Higgs invisible decay searches performed with $5,~20, ~36$ fb$^{-1}$ of 7, 8, and 13 TeV data provides the bound BR$(h\to{\rm{invisible}})\sim 0.19$ at 95$\%$ C.L. \cite{Sirunyan:2018owy}. The ATLAS bound translates into a bound on the Higgs coupling to DM $y_{hXX}\lesssim 0.01$ for $m_{X}\ll m_h$. This bound is shown in red in Fig.\,\ref{fig:relativecontributions2d}, and is seen to probe our freeze-in scenario at very low reheat temperatures $T_{RH}\sim$ GeV and for DM masses $\gtrsim$ 10 GeV. Future projections show that a bound on the Higgs invisible branching ratio at the level of $\sim 2\%$ can be achievable at the HL-LHC \cite{Cepeda:2019klc,deBlas:2019rxi}, which translates into $y_{hXX}\lesssim 4\times 10^{-3}$, slightly extending the coverage in parameter space. 

The phenomenology of the heavy scalar, $S$, depends not only on the coupling $y_{hXX}$ but also on its mixing with the SM Higgs boson, $\sin\theta_h$. It can be singly or pair produced at the LHC via its Higgs portal coupling, with the production cross section given by the corresponding cross section for a SM Higgs with the same mass, suppressed by  $\sin^2\theta_h$. Thus, non-negligible production at colliders requires this mixing to be sizable. For example, for the parameters used in Sec.\,\ref{Sec:sChannelScalar}, $m_S= 500$ GeV and $\sin\theta_h= 0.1$,  the heavy scalar production cross section calculated at NNLO+NNLL is $\sim 45$ fb \cite{LHCHiggsCrossSectionWorkingGroup:2016ypw}. This value of $\sin\theta_h$ will lead to prompt decays of $S$ into SM fermions and gauge bosons with width $\Gamma_S^{\rm{SM}}\sim 0.6$ GeV. Its decay width into dark matter particles depends on $y_{hXX}$ and $\sin\theta_h$. For the values needed to obtain the correct relic abundance via freeze-in, this width is generally negligible, except at very low values of $T_{\rm{RH}}\sim$ GeV and $m_X\gtrsim 10$ GeV (in and around the white region in Fig.\,\ref{fig:relativecontributions2d}, which corresponds to $y_{hXX}>1$). LHC searches for new scalar resonances do not yet constrain this heavy scalar with $m_S=500$ GeV, but it can be probed at the HL-LHC via its decays to $ZZ$ \cite{Cepeda:2019klc,CMS:2019qzn}. More exotic decays of $S$ are possible if the dark sector contains additional structure. This is a model dependent question, and a specific example that realizes  such possibilities is discussed in Appendix \ref{sspecific}.

\subsubsection{S-channel Vector Mediator}

\noindent\textbf{Kinetic Mixing}

 If the $Z-Z'$ kinetic mixing is significant, the $Z'$ can be produced copiously at the LHC, and its decays can provide observable signatures. 
 Depending on the parameters, the $Z'$ can decay dominantly into dark matter or to SM states. 
 
 The strongest LHC limits on $Z'$ resonances are derived from CMS and ATLAS searches for narrow dilepton resonances. The most important searches are: CMS search \cite{Sirunyan:2019wqq} for dimuon resonances in $110$ GeV $< m_{Z^{\prime}} < 200$ GeV; CMS search  \cite{CMS:2019tbu} and ATLAS search \cite{Aad:2019fac} for dilepton resonances at higher masses up to 6 TeV. Ref.\,\cite{Barnes:2020vsc} shows that in a kinetically mixed $Z'$ model where the $Z'$ decays 100\% into SM states, the bounds are at the level of $\epsilon\sim 10^{-2}$ across the range of masses we consider in our paper. In Fig.\,\ref{fig:contoursZ}, we set bounds on the $Z^\prime$ parameter space for two different values of $g_Dq_D=3\times 10^{-6},~3\times 10^{-5}$. For small values of $g_Dq_D$, these bounds are seen to be significant for low $T_{RH}$ models. 

Invisible decays of $Z'$ into dark matter, $Z'\to X\bar X$, could be probed, \textit{e.g.}~by monojet searches \cite{Aad:2021egl,Sirunyan:2017hci}. We have checked that the monojet cross sections predicted from the parameter space in Fig.\,\ref{fig:contoursZ} are several orders of magnitude smaller than what is currently probed by LHC searches. 

Indirect constraints from electroweak precision measurements \cite{Hook:2010tw,Curtin:2014cca} or measurements of $Z$ invisible decay width ($\Gamma_Z^{\rm{inv}}=499.0 \pm 1.5$ MeV \cite{Zyla:2020zbs}) are also very weak in the $m_{Z^\prime}\gtrsim 100$ GeV mass range that we focus on in this paper. 
 
 \bigskip
\noindent\textbf{\boldmath Gauged $L_\mu-L_\tau$}

The $Z^\prime$ gauge boson arising from gauging $L_\mu-L_\tau$ is only mildly constrained by collider data if  $m_{Z'}> m_Z$. The $Z^\prime$ can be produced at the LHC through its coupling to muons, taus, and neutrinos via the processes $pp\to Z^\prime\mu^+\mu^-$, $pp\to Z^\prime\nu\bar\nu$, and $pp\to Z^\prime\mu\nu_\mu$, where the muons can be replaced with taus.  
However, so far, LHC searches have only been performed in the mass range $(5-70)$ GeV, where the $Z^\prime$ is produced from $Z$ decay ($Z\to Z^\prime\mu^+\mu^-,Z^\prime\to\mu^+\mu^-$) \cite{Sirunyan:2018nnz}. 
Additional bounds can be obtained recasting ATLAS and CMS multilepton analyses. This has been done in \cite{Drees:2018hhs}, showing that the $Z^\prime$ masses up to 550 GeV are probed for $g^\prime=\mathcal O(1)$.

Additional constraints arise from high intensity experiments. The most stringent constraints come from the measurement of the neutrino trident process $\nu_\mu N\to \nu_\mu\mu^+\mu^- N$ \cite{Altmannshofer:2014cfa,Altmannshofer:2014pba} by the CCFR experiment \cite{Mishra:1991bv}, but this only constrains light $Z^\prime$ masses. Finally, the $L_\mu-L_\tau$ $Z^\prime$ can address the $(g-2)_\mu$ anomaly. However, for couplings $g^\prime\lesssim \mathcal O(1)$, this requires $m_{Z^\prime}\lesssim 200$ GeV \cite{Altmannshofer:2014cfa}.

All these bounds can in principle be affected by the mixing of the $Z^\prime$ with the SM hypercharge gauge boson. In the $L_\mu-L_\tau$ model, this mixing is generated at one loop (see Eq.\,(\ref{eq:mutaukinetic})). For the values of $g^\prime$ needed to produce the measured relic abundance, this mixing is very small and does not appreciably affect the collider bounds on $Z^\prime$.

\subsubsection{T-channel Mediator}
\label{sec:tcollider}

The t-channel mediator $S_T$ is a scalar with the same quantum numbers as the antifermion $\bar{f}$ ($f=t_R$ or $f=e_R$) that it couples to, 
whereas the $S_T-X$ system shares an effective $Z_2$-symmetry, as can be inferred from the Lagrangian in Eq.\,(\ref{eq:lagrangiantchannel}). Consequently, the $S_T$ phenomenology is very similar to that of a right-handed stop or slepton in the MSSM, with $X$ being the Bino LSP (lightest supersymmetry particle) and the $Z_2$-symmetry being the R-symmetry. 
The LHC searches for stops and sleptons pair production, followed by prompt decays into the corresponding fermion and the LSP (missing energy). The range of values of the coupling $y_T$ needed to obtain the correct dark matter abundance results in $S_T$ decays that are prompt for the purpose of LHC searches. Therefore, the most stringent bounds on $S_T$ 
arise from the ATLAS and CMS searches for promptly decaying stops \cite{Aad:2020sgw,Aad:2020aob,Sirunyan:2019glc,Sirunyan:2021mrs} and sleptons \cite{Aad:2019vnb,Sirunyan:2020eab}. The LHC bounds on such particles, with an essentially massless LSP (recall that in Sec.\,\ref{sec:tchannel}, we focused on benchmarks with $m_X=1$ GeV), are approximately $m_{S_T}\! \gtrsim \!1200$ GeV ($\gtrsim\!400$ GeV) for $S_T$ coupled to the top quark (electron), based on $\sim 140$/fb LHC Run II data. 

\subsection{Direct Detection}
\label{subsec:directdetection}
Direct detection signals are generically suppressed in dark matter freeze-in models due to the feeble couplings involved. However, as we discuss here, our low reheat temperature freeze-in scenarios have better direct detection prospects due to generically larger SM-dark matter interactions.

For s-channel Higgs mediated models, the direct detection spin-independent DM-nucleon cross section can be approximated by 
\beq
\sigma_{SI}^{n,h}\simeq 7\times 10^{-43} \left(\frac{\mu_{Xn}}{\rm{GeV}}\right)^2 (y_{hXX}\cos\theta_h)^2 \text{cm}^2\,,
\label{eq:ddhiggs}
\eeq
where we have neglected the contribution of the heavy scalar, and $\mu_{Xn}$ is the dark matter-nucleon reduced mass. 
The XENON1T result \cite{XENON:2018voc} constraints some parts of the parameter space of our freeze-in framework for dark matter masses above $\sim\!10$ GeV (see the orange curve in Fig.\,\ref{fig:relativecontributions2d}). 

For Dirac dark matter in the kinetically mixed scenario, the spin-independent scattering cross section receives contributions from both the $Z$ and the $Z^\prime$ and is given by \be\label{eq:sigmaDDvector}
\sigma_{SI}^{n,Z}
\simeq 3\times 10^{-38} \left(\frac{\mu_{Xn}}{\rm{GeV}}\right)^2 q_D^2 g_D^2\epsilon^2 (m_Z/m_Z')^4 \,\text{cm}^2\,.
\ee
For light dark matter masses as we consider in the plots of Sec.\,\ref{Sec:sChannelGauge}, the most relevant bounds are from the CRESST-III experiment \cite{CRESST:2019jnq}, which constrain $\sigma_{SI}\sim 10^{-37}$ cm$^{2}$.
Future experiments such as SuperCDMS and NEWS-G will improve on this by several orders of magnitude \cite{Bondarenko:2019vrb}. However, even these more stringent projections are unable to probe the parameter space relevant for the freeze-in scenario discussed in this paper. For larger values of the dark matter mass and relatively large values of $\epsilon g_Dq_D$, the cross sections in Eq.\,(\ref{eq:sigmaDDvector}) could be tested with XENON1T data \cite{XENON:2018voc} or with future LZ \cite{LZ:2015kxe} or DARWIN \cite{Schumann:2015cpa} data.

For the t-channel mediator framework, different processes can play the leading role for direct detection. For the model with $S_T$ coupled to right-handed electrons, dark matter can scatter with electrons at tree-level (s-channel analog of Fig.\,\ref{fig:Tdiagrams}\,(a)) with an approximate cross section  
\beq
\sigma_e=\frac{y_T^4~\mu_{X\,e}^2}{m_{S_T}^4}\approx 10^{-41}\,y_T^4\left(\frac{100\,\text{GeV}}{m_{S_T}}\right)^4 \text{cm}^2.
\label{eq:ddZ}
\eeq
For $y_T\sim 10^{-5}$ as roughly needed for the correct relic abundance, this cross section is far too small to be probed experimentally. Scattering with nuclei are mediated by one-loop penguin diagrams (analogous to Fig.\,\ref{fig:Tdiagrams}\,(d,e)) with the $Z/h$ mediating the scattering with nuclei. For the model with $S_T$ coupled to right-handed top quarks, analogous diagrams mediated by gluons are also relevant.  Ref.\,\cite{Herrero-Garcia:2018koq} finds that the direct detection cross section mediated by gauge bosons features a further suppression by $|q|^2/m_f^2$, where $q$ is the momentum transferred in the scattering process. The Higgs penguin is also negligible since the $h X\bar{X}$ effective vertex is much smaller than the $Z X\bar{X}$ effective vertex due to additional mass suppressions (see the discussion below Eq.\,(\ref{eq:hloopwidth})). Therefore, it will be quite challenging to probe the t-channel mediator framework considered in this paper at direct detection experiments.

\section{Summary}
\label{sec:discussion}

In this paper, we have studied scenarios of dark matter freeze-in where the mediator particle that gives rise to interactions between the dark matter ($X$) and the Standard Model is heavier than the reheat temperature, $T_{RH}$, in the early Universe. In such setup, the standard approach is to integrate out the mediator and focus on an effective field theory (EFT) with higher dimensional interactions between the SM and DM. 
We examined the validity of this approach in the regime $M_{med}\gtrsim T_{RH}$.

We studied three classes of s-channel mediator frameworks: (i) a heavy scalar that mixes with the SM Higgs boson, (ii) a heavy $Z'$ that mixes kinetically with the SM hypercharge, and (iii) a $Z'$ gauge boson from a gauged $L_\mu-L_\tau$ symmetry that couples directly to both SM particles and dark matter. In all cases, the EFT approach (integrating out the mediators -- including the SM Higgs / $Z$ boson -- and focusing only on the resulting $f\bar{f}\to X\bar{X}$ processes) captures the correct dark matter freeze-in abundance at very low $T_{RH}$, where dark matter is dominantly produced through non-resonant annihilation of SM fermions: $T_{RH}\lesssim 5$ GeV for the scalar case, $T_{RH}\lesssim 7$ GeV for the kinetically mixed $Z'$ case, and $T_{RH}\lesssim 0.025\, m_{Z'}$ for the $L_\mu-L_\tau$ $Z'$ model. However, at higher $T_{RH}$, the resonant enhancement of the s-channel annihilation cross section, not captured within the EFT, becomes important. The predicted dark matter freeze-in abundance from fermion annihilation in the EFT can deviate from the full result by several orders of magnitude (see Fig.\,\ref{fig:ratios}). In this regime, the dark matter abundance is instead appropriately captured by considering decays of exponentially suppressed thermal abundances of the mediators in the bath. 

Similarly, we studied t-channel mediator scenarios with couplings to top quarks or electrons (both right-handed). For the $e_R$ case, we found that the EFT calculation reproduces the correct dark matter abundance for $T_{RH}\lesssim 0.05 \,M_{med}$. At higher $T_{RH}$, dark matter is dominantly produced through coannihilation processes $(Z/\gamma)\,e\to S_T^\dagger\,X$. Similarly, for the $t_R$ case, $g t\to S_T^\dagger\,X$ dominates at high temperatures, $T_{RH}\gtrsim 0.04 \,M_{med}$. In contrast, for low $T_{RH}$ scenarios, given the suppressed abundance of top quarks in the thermal bath, loop decays of the SM $Z$ boson induced by the t-channel mediator are the dominant source of dark matter abundance. 

We thus find that in both s- and t-channel scenarios, novel channels that are not captured by the EFT dominate dark matter production even when $T_{RH}$ is more than an order of magnitude below the mass of the mediator. It is therefore important to include such contributions when studying dark matter freeze-in production with heavy mediators. Furthermore, these new production channels also change the momentum distribution of dark matter, peaking it towards a warmer distribution than standard freeze-in scenarios. 

Finally, we discussed the collider phenomenology of the heavy mediators and the prospects of testing these scenarios at direct detection experiments. We find that, in contrast to mediators in standard freeze-in scenarios, the mediators in our setup can have large couplings with both DM and SM particles, leading to qualitatively different collider phenomenology compared to standard freeze-in setups. Parts of the parameter space of our low $T_{RH}$ scenarios are already probed by LHC searches for Higgs invisible decays, searches for prompt dilepton resonances and for SUSY stops and sleptons, as well as by dark matter direct detection experiments.

\section*{Acknowledgements}
We thank Wolfgang Altmannshofer, Jeff Dror, Aaron Pierce, and Joshua Ruderman for helpful discussions. We are especially grateful to Hiren Patel for several discussions on the use of Package-X. The research of SG is supported in part by the National Science Foundation CAREER grant PHY- 1915852 and by the National Science Foundation under Grant No. NSF PHY-1748958. SG thanks the Aspen Center for Physics for hospitality when part of this work was done. The Aspen Center for Physics is supported by National Science Foundation grant PHY-1607611. The work of BS is supported by the Deutsche Forschungsgemeinschaft under Germany’s Excellence Strategy - EXC 2121 Quantum Universe - 390833306.


\appendix 

\section{Contributions from the Epoch Before Radiation Domination}
\label{app:inflaton}

In this Appendix, we consider contributions from the era before radiation domination, when the Universe was governed by the energy density of the inflaton, $\phi$, when the temperature of the radiation bath could have been larger than $T_{RH}$ \cite{Chung:1998rq,Giudice:2000ex}. Since the annihilation and decay freeze-in contributions discussed in this paper become more efficient at higher temperatures, this earlier epoch, despite being very short in duration, can contribute non-negligibly to the current dark matter relic density. 

The transfer of the inflaton energy density, $\rho_\phi$, into the thermal (radiation) bath energy density, $\rho_R$, is governed by the following differential equations:
\beq
\dot{\rho}_\phi=-3H\rho_\phi-\Gamma_\phi\rho_\phi,~~~~\dot{\rho}_R=-4H\rho_R+\Gamma_\phi\rho_\phi\,,
\eeq
where we assume that $\phi$ decays into radiation with a rate $\Gamma_\phi$. During this evolution, the temperature of the radiation bath is given by
\beq
T=\left(\frac{30}{g_*(T)\pi^2}\,\rho_R\right)^{1/4}\,,
\eeq
where $g_*(T)$ is the effective number of degrees of freedom in the bath at temperature $T$. If we assume an instantaneous decay of the inflaton energy density, the conventional reheat temperature signifying the temperature of the Universe at the onset of radiation domination can be written as
\beq
T_{RH}=\left(\frac{90}{8\pi^3 g_*(T_{RH})}\right)^{1/4}\sqrt{\Gamma_\phi M_{Pl}}\,.
\eeq
The maximum temperature the radiation bath reaches during this phase can be written as
\beq
\frac{T_{MAX}}{T_{RH}}\approx \left(\frac{\rho_{\phi0}}{\Gamma_\phi^2 M_{Pl}^2}\right)^{1/8}\, ,
\eeq
where $\rho_{\phi0}$ is the initial energy density of the inflaton. Thus, temperatures prior to the radiation dominated era can be higher than $T_{RH}$ if $\rho_{\phi 0}> \Gamma_\phi^2 M_{Pl}^2$, i.e., if the initial inflaton energy density is large and transferred to the radiation bath at a very slow rate. 
Parameterizing 
\beq
\rho_{\phi0}=M_\phi^4,~~~~\Gamma_\phi=\alpha_\phi M_\phi,
\eeq
we expect $T_{MAX}>T_{RH}$ if $\alpha_\phi < M_\phi/M_{Pl}$. In this case, the higher temperatures during this era can result in greater dark matter production from decays of the heavy mediator(s), whose abundances are no longer Boltzmann suppressed, as well as annihilation processes, which can proceed faster. We numerically solve the above differential equations and calculate the production of dark matter from various processes, and find, indeed, that the dark matter abundance from this era dominates over the abundance from the subsequent radiation dominated era for $\alpha_\phi \lesssim M_\phi/M_{Pl}$ for which $T_{MAX}>T_{RH}$, as discussed above. Thus, for this epoch before radiation domination to contribute negligibly to the dark matter abundance in the Universe, we must assume $\alpha_\phi\gg M_\phi/M_{Pl}$.

\section{Specific Models}\label{app:models}

In this Appendix, we discuss some specific realizations of the simplified models discussed in this paper, and explore how model-specific details can affect phenomenological aspects.

\subsection{S-channel Mediator: Sterile Neutrino Dark Matter}
\label{sspecific}

Here we discuss a specific model for the s-channel simplified model presented in Sec. \ref{Sec:sChannelScalar}. We consider an extension of the canonical (type I) seesaw mechanism, where singlet right-handed neutrinos, $N_j$, act as portals to a hidden sector (see e.g. \cite{Shakya:2018qzg,Roland:2014vba,Roland:2015yoa,Shakya:2015xnx,Roland:2016gli,Shakya:2016oxf,Patel:2019zky} for details):
\be
\mathcal{L}\supset y_{ij} L_i h N_j+M_j \bar{N}^c_j N_j + y'_{kj}L'_kh'N_j.
\label{eq:lagrangian1}
\ee
Here $L_i$ ($i=1,2,3$) and $h$ are the SM lepton doublet and Higgs fields, respectively, $y_{ij},y'_{kj}$ are dimensionless Yukawa couplings, and $L'_k,\,h'$ are hidden sector fermion and scalar states, charged under a hidden sector $U(1)'$ symmetry. Integrating out the $N_j$, which we assume to be much heavier than other scales in the model (these will henceforth be ignored, and the notation $N_k$ will refer to the light sterile states $L'$) and dropping indices for simplicity, we have: 
\be
\mathcal{L}\supset \frac{1}{M}y^2 (L h)^2 +\frac{1}{M}y y' (L h)(L'h')+\frac{1}{M}y'^2 (L'h')^2.
\label{eq:lagrangian4}
\ee 

As before, we also assume a quartic term $\lambda\,h'^2 h^2$ that leads to mixing between the two scalars. Here, the scalar $h'$ serves as the heavy mediator, $S$, while one of the sterile neutrinos, \textit{e.g.} $N_1$, is dark matter. If the hidden sector scalar obtains a VEV, $v'$, the dark matter mass is given by $y'^2 v'^2/M$ and is naturally much smaller than the mediator mass $m_{S}\sim v'$, if $M\gg v'$. If the $U(1)'$ is gauged, this model also includes the vector ($Z'$) mediator with mass $m_{Z'}\sim g' v'$, which is again significantly heavier than dark matter.

The effective Higgs-sterile neutrino couplings are approximately
\bea
y_{h\nu N_j}&=&\frac{y y' v'}{M},~~~
y_{hN_iN_j}=2\sin\theta_h \frac{y'^2 v'}{M}\approx2\frac{\lambda v y'^2 v'^2}{m_h'^2 M}\approx2\frac{\lambda v \,\sqrt{m_{N_i}m_{N_j}}}{m_h'^2}\,,\\
y_{h'N_iN_j}&=&2\cos\theta_h \frac{y'^2 v'}{M}\approx 2\cos\theta_h \frac{\sqrt{m_{N_i}m_{N_j}}}{v'}\,.\nonumber\label{eq:Coupl}
\eea
Here, $h, h'$ denote the SM-like and heavy Higgs mass eigenstates, and $\nu, N$ denote the SM and hidden sector sterile neutrinos. Note that all of the couplings are suppressed by the heavy scale, $M$, and are therefore expected to be small.  

Considering a specific model also gives rise to a novel phenomenology that is not captured by the simplified model discussion. While freeze-in production of dark matter $N_1$ proceeds via both annihilation of SM fermions and decays of $h,h'$, in this specific model there are also additional production channels, due to the presence of the additional sterile neutrinos $N_i$. These can lead to both annihilation $N_iN_i\to N_1 N_1$ and decay $N_i\to N_1 N_1 \nu$ contributions via the Higgs and neutrino portals. While the decays of $h, h'$ to $N_1$ are invisible, the presence of the additional sterile neutrino states also gives rise to new collider signatures: $h, h'\to N_i N_i$, which can then further decay into SM fermions, e.g. as $N_i\to \nu e^+e^-$ with displaced vertices (see e.g. \cite{Nemevsek:2016enw}).

\subsection{T-channel Mediator: Axino Dark Matter}

Here we discuss a specific model for the t-channel simplified model presented in Sec. \ref{sec:tchannel}. The model contains axino dark matter and is based on Ref.\,\cite{Covi:2002vw} (the interested reader is referred to this paper for more details). Models that solve the strong CP problem using the Peccei Quinn (PQ) symmetry contain a new particle, the axion. Supersymmetric extensions also contain its superpartner, the axino. The couplings of the axino to the MSSM particles (as is the case for the axion) are suppressed by the PQ scale, $f_a$. Therefore, the axino only has feeble couplings to the remainder of the MSSM field content and, as we discuss below, can be a dark matter candidate with an abundance set by the freeze-in mechanism.  

Since the axino is part of a chiral multiplet, it does not acquire a tree-level Majorana soft mass term from supersymmetry breaking, but can obtain a small effective mass due to mixing with neutral heavy states by virtue of the presence of Dirac mass terms, or from loop effects. The axino can therefore naturally be much lighter than the MSSM particles. We henceforth treat the axino mass as a free parameter. We also confine ourselves to scenarios where $T_{RH}$ is significantly below the scales of both PQ and supersymmetry breaking, so that the axino is the only relevant SUSY particle in the early Universe. 

Following Ref.\,\cite{Covi:2002vw}, we focus on the KSVZ-type axion models \cite{Kim:1979if,Shifman:1979if}, where SM particles are not charged under the $U(1)_{PQ}$ symmetry. In this setup, the axino couples to MSSM states via loops of heavy PQ states (with masses at the PQ scale, $f_a$). The Lagrangian contains a dimension-5 axino-gluon-gluino vertex, which is given by 
\beq
\mathcal{L}_{\tilde{a} g \tilde{g}}=  \frac{\alpha_S}{8 \pi f_a}\bar{\tilde{a}}\gamma_5 \sigma^{\mu\nu} \tilde{g}^b G^b_{\mu\nu}.
\eeq

Likewise, an axino-quark-squark vertex arises at two loops and is given by 
  \cite{Covi:2002vw}
\beq
g_{eff}\approx  \frac{\alpha_S^2}{\sqrt{2} \pi^2}\frac{m_{\tilde{g}}}{f_a} \text{log}\left(\frac{f_a}{m_{\tilde{g}}}\right)\,.
\label{eq:axinoquarksquark}
\eeq

Depending on the mass hierarchy between the squarks and gluinos, either of these two interactions can dominate dark matter production in the early Universe through the t-channel processes $gg\to aa, q\bar q\to aa$. If we assume that gluinos are much heavier than squarks and can be neglected, then the only relevant interaction for dark matter production is the axino-quark-squark interaction. Thus, this well motivated supersymmetric framework maps onto our t-channel simplified model Lagrangian in Eq.\,(\ref{eq:lagrangiantchannel}), with the squarks acting as the heavy mediator, $S_T$, and $g_{eff}$ ($\ll\!1$, as $m_{\tilde{g}}\ll f_a$) as the coupling $y_T$.

\bibliographystyle{utphys}
\bibliography{freeze_in_heavy_mediator}

\end{document}